%% file: TijHumOpFuz.tex
\documentclass[11pt]{article}
\usepackage{graphicx,amsmath,amssymb,amsfonts,times,hyperref}
\input{tcilatex}

\begin{document}

\title{Fuzzy Control Strategies in Human Operator and Sport Modeling}
\author{Tijana Ivancevic\thanks{Society for Nonlinear Dynamics in Human Factors, Adelaide, Australia}, ~Bojan Jovanovic, and Sasa Markovic\thanks{Faulty of Sport Sciences, University of Nis, Serbia}}
\date{}
\maketitle\vspace{3cm}

\abstract{The motivation behind mathematically modeling the
\textit{human operator} is to help explain the response
characteristics of the complex dynamical system including the
human manual controller. In this paper, we present two different fuzzy
logic strategies for human operator and sport modeling: fixed fuzzy--logic inference control and adaptive fuzzy--logic control, including neuro--fuzzy--fractal control. As an application of the presented fuzzy strategies, we present a fuzzy-control based tennis simulator.}\newpage

\tableofcontents

\section{Introduction}

Despite the increasing trend toward automation, robotics and
artificial intelligence (AI) in many environments, the \emph{human
operator} will probably continue for some time to be integrally
involved in the control and regulation of various machines (e.g.,
missile--launchers, ground vehicles, watercrafts, submarines,
spacecrafts, helicopters, jet fighters, etc.). A typical manual
control task is
the task in which control of these machines is accomplished by \textit{%
manipulation of the hands or fingers} \cite{Wickens}. As
human--computer interfaces evolve, interaction techniques
increasingly involve a much more continuous form of interaction
with the user, over both human--to--computer (input) and
computer--to--human (output) channels. Such interaction could
involve gestures, speech and animation in addition to more
`conventional' interaction via mouse, joystick and keyboard. This
poses a problem for the design of interactive systems as it
becomes increasingly necessary to consider interactions occurring
over an interval, in continuous time.

The so--called \textit{manual control theory} developed out of the efforts
of feedback control engineers during and after the World War II, who
required models of human performance for continuous military tasks, such as
tracking with anti--aircraft guns \cite{Wiener}. This seems to be an area
worth exploring, firstly since it is generally concerned with systems which
are controlled in continuous time by the user, although discrete time
analogues of the various models exist. Secondly, it is an approach which
models both system and user and hence is compatible with research efforts on
`synthetic' models, in which aspects of both system and user are specified
within the same framework. Thirdly, it is an approach where continuous
mathematics is used to describe functions of time. Finally, it is a theory
which has been validated with respect to experimental data and applied
extensively within the military domains such as avionics.

The premise of manual control theory is that for certain tasks, the
performance of the human operator can be well approximated by a describing
function, much as an inanimate controller would be. Hence, in the literature
frequency domain representations of behavior in continuous time are applied.
Two of the main classes of system modelled by the theory are \textit{%
compensatory} and \textit{pursuit} systems. A system where only the error
signal is available to the human operator is a compensatory system. A system
where both the target and current output are available is called a pursuit
system. In many pursuit systems the user can also see a portion of the input
in advance; such tasks are called \textit{preview tasks} \cite{doh}.

A simple and widely used model is the `crossover model' \cite{2},
which has two main parameters, a \textit{gain} $K$ and a
\textit{time delay} $\tau$, given by the transfer function in the
Laplace transform $s$ domain
$$H=K{\mathrm{e}^{-\tau s}\over s}.
$$
Even with this simple model we can investigate some quite
interesting phenomena. For example consider a compensatory system
with a certain delay, if we have a low gain, then the system will
move only slowly towards the target, and hence will seem sluggish.
An expanded version of the crossover model is given by the
transfer function \cite{Wickens}
$$ H=K\frac{(T_Ls+1)\,\mathrm{e}^{-(\tau s+\alpha/s)}}{(T_Is+1)(T_Ns+1)},
$$
where $T_L$ and $T_I$ are the lead and lag constants (which
describe the \textit{equalization} of the human operator), while
the first--order lag $(T_NS+1)$ approximates the neuromuscular lag
of the hand and arm. The expanded term $\alpha/s$ in the time
delay accounts for the `phase drop', i.e., increased lags observed
at very low frequency \cite{GaneshSprBig}.

Alternatively if the gain $K$ is very high, then the system is
very likely to overshoot the target, requiring an adjustment in
the opposite direction, which may in turn overshoot, and so on.
This is known as `oscillatory behavior'. Many more detailed models
have also been developed; there are `anthropomorphic models',
which have a cognitive or physiological basis. For example the
`structural model' attempts to reflect the structure of the human,
with central nervous system, neuromuscular and vestibular
components \cite{doh}. Alternatively there is the `optimal control
modeling' approach, where algorithmic models which very closely
match empirical data are used, but which do not have any direct
relationship or explanation in terms of human neural and cognitive
architecture \cite{3}. In this model, an operator is assumed to
perceive a vector of displayed quantities and must exercise
control to minimize a \textit{cost functional} given by
\cite{Wickens}
$$J=E\{\lim_{T\to\infty} \frac{1}{T} \int^T_0 [q_iy_i^2(t) + \sum_i(r_iu^2(t)
+ g_i\dot{u}^2(t))]dt \},
$$
which means that the operator will attempt to minimize  the
expected value $E$ of some weighted combination of squared display
error $y$, squared control displacement $u$ and squared control
velocity $\dot u$. The relative values of the weighting constants
$q_i, r_i, g_i$ will depend upon the relative importance of
control precision, control effort and fuel expenditure.

In the case of manual control of a vehicle, this modeling yields
the `closed--loop' or `operator--vehicle' dynamics. A quantitative
explanation of this closed--loop behavior is necessary to
summarize operator behavioral data, to understand operator control
actions, and to predict the operator--vehicle dynamic
characteristics. For these reasons, control engineering
methodologies are applied to modeling human operators. These
`control theoretic' models primarily attempt to represent the
operator's control behavior, not the physiological and
psychological structure of the operator \cite{0,00}. These models
`gain in acceptability' if they can identify features of these
structures, `although they cannot be rejected' for failing to do
so \cite{1}.

One broad division of human operator models is whether they
simulated a continuous or discontinuous operator control strategy.
Significant success has been achieved in modeling human operators
performing compensatory and pursuit tracking tasks by employing
continuous, quasi--linear operator models. Examples of these
include the crossover optimal control models mentioned above.

Discontinuous input behavior is often observed during manual control of
large amplitude and acquisition tasks \cite{2,4,5,6}. These discontinuous
human operator responses are usually associated with \textit{precognitive
human control behavior} \cite{2,7}. Discontinuous control strategies have
been previously described by `bang--bang' or relay control techniques. In
\cite{8}, the authors highlighted operator's preference for this type of
relay control strategy in a study that compared controlling high--order
system plants with a proportional verses a relay control stick. By allowing
the operator to generate a sharper step input, the relay control stick
improved the operators' performance by up to 50 percent. These authors
hypothesized that when a human controls a high--order plant, the operator
must consider the error of the system to be dependent upon the integral of
the control input. Pulse and step inputs would reduce the integration
requirements on the operator and should make the system error response more
predictable to the operator.

Although operators may employ a \emph{bang--bang control} strategy, they often
impose an internal limit on the magnitude of control inputs. This internal
limit is typically less than the full control authority available \cite{2}.
Some authors \cite{9} hypothesized that this behavior is due to the
operator's recognition of their own reaction time delay. The operator must
tradeoff the cost of a switching time error with the cost of limiting the
velocity of the output to a value less than the maximum.

A significant amount of research during the 1960's and 1970's examined
discontinuous input behavior by human operators and developed models to
emulate it \cite{7,10,11,12,13,14,15,16,17}. Good summaries of these efforts
can be found in \cite{18}, \cite{4}, \cite{2} and \cite{0,00}. All of these
efforts employed some type of \textit{relay} element to model the
discontinuous input behavior. During the 1980's and 1990's, pilot models
were developed that included switching or discrete changes in pilot behavior
\cite{19,20,21,22,5,6}.

Recently, the so-called `variable structure control' techniques were applied
to model human operator behavior during acquisition tasks \cite{0,00}. The
result was a coupled, multi--input model replicating the discontinuous
control strategy. In this formulation, a switching surface was the
mathematical representation of the human operator's control strategy. The
performance of the variable strategy model was evaluated by considering the
longitudinal control of an aircraft during the visual landing task.

For a review of classical feedback control theory in the context of human operator modelling
see \cite{GaneshSprSml,GaneshSprBig,GCompl} and contrast
it with nonlinear and stochastic dynamics (see \cite{TacaNODY,StrAttr,Complexity}). For similar approaches to sport modelling, see \cite{TacaSpr}.

In this paper, we present two different fuzzy
logic strategies for human operator and sport modeling: fixed fuzzy--logic inference control and adaptive fuzzy--logic control, including neuro--fuzzy--fractal control. As an application of the presented fuzzy strategies, we present a fuzzy-control based tennis simulator.\bigbreak\bigbreak

\section{Fixed Fuzzy Control in Human Operator Modeling}

Modeling is the name of the game in any intelligence, be it human
or machine. With the model and its exercising we can look forward
in time with predictions and prescriptions and backward in time
with diagnostics and explanations. With these time binding
information structures we can make decisions and estimations in
the here and now for purposes of efficiency, efficacy and control
into the future. We and our machines hope to look into the future
and the past so we may act intelligently now.\footnote{The Fuzzy
Cognitive Map, Fuzzy Systems Engineering.}

Recall that fuzzy logic is a departure from classical two--valued
sets and logic, that uses `soft' linguistic (e.g. large, hot,
tall) system variables and a continuous range of truth values in
the interval [0,1], rather than strict binary (True or False)
decisions and assignments.

Formally, fuzzy logic is a structured, model--free estimator that
approximates a function through linguistic input/output
associations.

Fuzzy rule-based systems apply these methods to solve many types
of `real--world' problems, especially where a system is difficult
to model, is controlled by a human operator or expert, or where
ambiguity or vagueness is common. A typical fuzzy system consists
of a rule base, membership functions, and an inference
procedure.\bigbreak

The key \textit{benefits} of fuzzy logic design are:
\begin{enumerate}
    \item Simplified \& reduced development cycle;
    \item Ease of implementation;
    \item Can provide more `user--friendly' and efficient
    performance;
\end{enumerate}

Some fuzzy logic \textit{applications} include:\begin{enumerate}
    \item Control (Robotics, Automation, Tracking, Consumer
    Electronics);
    \item Information Systems (DBMS, Info. Retrieval);
    \item Pattern Recognition (Image Processing, Machine Vision);
    \item Decision Support (Adaptive HMI, Sensor Fusion).
\end{enumerate}

Recall that conventional controllers are derived from control
theory techniques based on mathematical models of the open--loop
process, called system, to be controlled. On the other hand, in a
fuzzy logic controller, the dynamic behavior of a fuzzy system is
characterized by a set of linguistic description rules based on
expert knowledge. The expert knowledge is usually of the form:

IF (a set of conditions are satisfied) THEN (a set of consequences
can be inferred).

Since the \textit{antecedents} and the \textit{consequents} of
these IF--THEN rules are associated with fuzzy concepts
(linguistic terms), they are often called \textit{fuzzy
conditional statements}. In this terminology, a fuzzy control rule
is a fuzzy conditional statement in which the antecedent is a
condition in its application domain and the consequent is a
control action for the system under control. Basically, fuzzy
control rules provide a convenient way for expressing control
policy and domain knowledge. Furthermore, several linguistic
variables might be involved in the antecedents and the conclusions
of these rules.

Furthermore, several linguistic variables might be involved in the
antecedents and the conclusions of these rules. When this is the
case, the system will be referred to as a multi--input
multi--output fuzzy system.

The most famous fuzzy control application is the subway car
controller used in Sendai (Japan), which has outperformed both
human operators and conventional automated controllers.
Conventional controllers start or stop a train by reacting to
position markers that show how far the vehicle is from a station.
Because the controllers are rigidly programmed, the ride may be
jerky: the automated controller will apply the same brake pressure
when a train is, say, 100 meters from a station, even if the train
is going uphill or downhill.

In the mid-1980s engineers from Hitachi used fuzzy rules to
accelerate, slow and brake the subway trains more smoothly than
could a deft human operator. The rules encompassed a broad range
of variables about the ongoing performance of the train, such as
how frequently and by how much its speed changed and how close the
actual speed was to the maximum speed. In simulated tests the
fuzzy controller beat an automated version on measures of riders'
comfort, shortened riding times and even achieved a 10 percent
reduction in the train's energy consumption \cite{KoskoIsaka}.

\subsection{Fuzzy Inference Engine}

Recall that a crisp (i.e., ordinary mathematical) set $X$ is
defined by a binary characteristic function $\mu _{X}(x)$ of its
elements $x$
\begin{equation*}
\mu _{X}(x)=\left\{
\begin{array}{c}
1,\qquad \text{if \ }x\in X, \\
0,\qquad \text{if \ }x\notin X,
\end{array}
\right.
\end{equation*}
while a fuzzy set is defined by a continuous characteristic
function
\begin{equation*}
\mu _{X}(x)=\left[ 0,1\right],
\end{equation*}
including all (possible) real values between the two crisp extremes $1$ and $%
0$, and including them as special cases.

A fuzzy set set $X$ is a collection of ordered pairs

\begin{equation}
X=\{\left( x,\mu (x)\right) \},  \label{f_set}
\end{equation}%
where $\mu (x)$ is the \textit{membership function} representing
the grade of membership of the element $x$ in the set $X$. A
single pair is called a fuzzy \textit{singleton}.

\begin{figure}[tbh]
\centerline{\includegraphics[width=12cm]{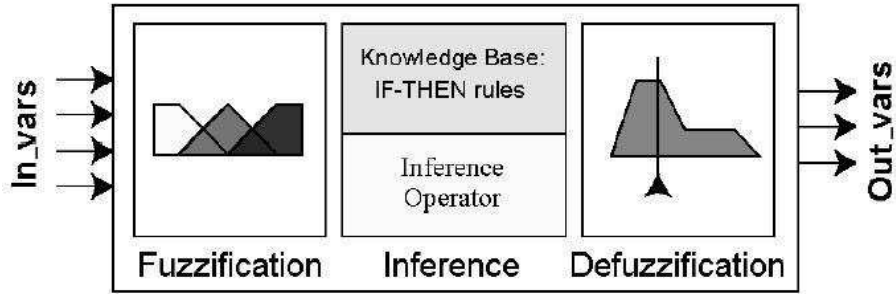}}
\caption{Basic structure of the fuzzy inference engine.}
\label{FuzzyInferenceEngine}
\end{figure}

Like neural networks, the fuzzy logic systems are generic
\textit{nonlinear function approximators} \cite{Kosko}. In the
realm of fuzzy logic this generic nonlinear function approximation
is performed by means of fuzzy inference engine. The \textit{fuzzy
inference engine} is an \textit{input--output dynamical system}
which \textit{maps} a set of input linguistic variables
($IF-$part) into a set of output linguistic variables
($THEN-$part). It has three sequential modules (see Figure
\ref{FuzzyInferenceEngine}):
\begin{enumerate}
    \item \emph{Fuzzification}\index{fuzzification}; in this module numerical crisp input
variables are fuzzified; this is performed as an overlapping
partition of their universes of discourse by means of fuzzy
membership functions $\mu (x)$ (\ref{f_set}), which can have
various shapes, like triangular, trapezoidal, Gaussian--bell,
\begin{equation*}
\mu (x)=\exp \left[ \frac{-(x-m)^{2}}{2\sigma ^{2}}\right] ,
\end{equation*}%
(with mean $m$ and standard deviation $\sigma $), sigmoid
\begin{equation*}
\mu (x)=\left[ 1+\left( \frac{x-m}{\sigma }\right) ^{2}\right]
^{-1},
\end{equation*}%
or some other shapes (see Figure \ref{UnivDiscour}).

\begin{figure}[tbh]
\centerline{\includegraphics[width=8cm]{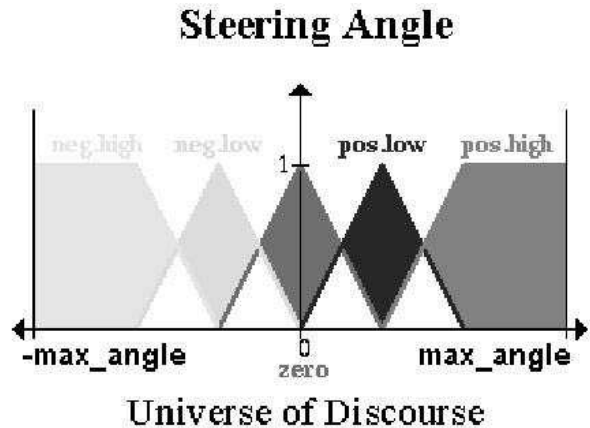}}
\caption{Fuzzification example: set of triangular--trapezoidal
membership functions partitioning the universe of discourse for
the angle of the hypothetical steering wheel; notice the white
overlapping triangles.} \label{UnivDiscour}
\end{figure}

B. Kosko and his students have done extensive computer simulations
looking for the best shape of fuzzy sets to model a known test
system as closely as possible. They let fuzzy sets of all shapes
and sizes compete against each other. They also let neural systems
tune the fuzzy--set curves to improve how well they model the test
system. The main conclusion from these experiments is that
`triangles never do well' in such contests. Suppose we want an
adaptive fuzzy system $F:\Bbb{R}^{n}\rightarrow \Bbb{R}$ to
approximate a test function or approximand
$f:\Bbb{R}^{n}\rightarrow \Bbb{R}$ as closely as possible in the
sense of minimizing the mean--squared error between them, $\left(
\left\| f-F\right\| ^{2}\right) $. Then the $i$th scalar `sinc'
function (as commonly used in signal processing),
\begin{equation}
\mu_{i}(x)=\frac{\sin \left( \frac{x-m_{i}}{d_{i}}\right) }{\frac{x-m_{i}}{%
d_{i}}},\qquad  i=1,...,n,  \label{sinc}
\end{equation}
with center $m_{i}$ and dispersion (width) $d_{i}=\sigma_i^2>0$,
often gives the best performance for $IF-$part mean--squared
function approximation, even though this generalized function can
take on negative values (see \cite{Kosko3}).
    \item \emph{Inference}\index{inference}; this module has two submodules:

(i) The expert--knowledge base consisting of a set of $IF-THEN$
rules relating input and output variables, and

(ii) The inference method, or implication operator, that actually
combines the rules to give the fuzzy output; the most common is
\textit{Mamdani Min--Max inference}, in which the membership
functions for input variables are first combined inside the
$IF-THEN$ rules using $AND$ ($\cap $, or $Min$) operator, and then
the output fuzzy sets from different $IF-THEN$ rules are combined
using $OR$ ($\cup $, or $Max$) operator to get the common fuzzy
output (see Figure \ref{Mamdani}).

\begin{figure}[tbh]
\centerline{\includegraphics[width=12cm]{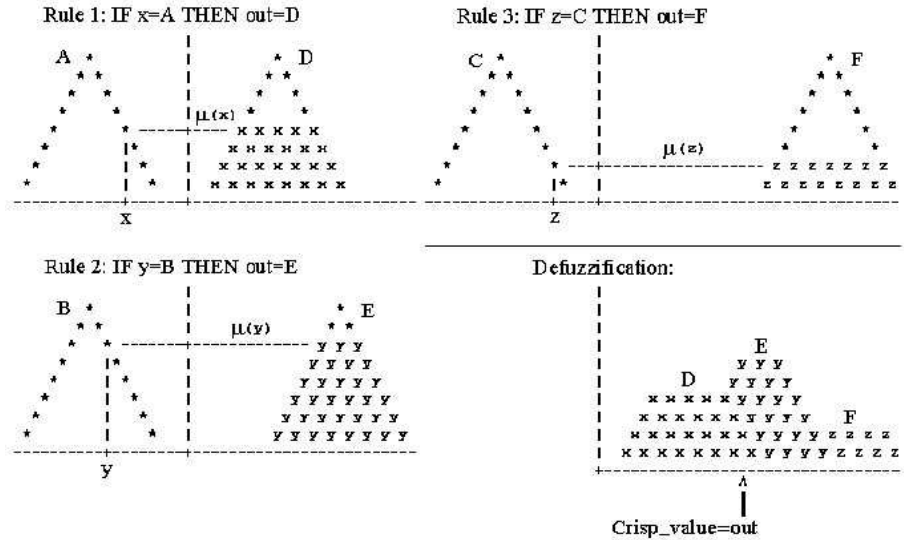}}
\caption{Mamdani's Min--Max inference method and Center of Gravity
defuzzification.} \label{Mamdani}
\end{figure}
    \item \emph{Defuzzification}\index{defuzzification}; in this module fuzzy outputs from the
inference module are converted to numerical crisp values; this is
achieved by one of the several defuzzification algorithms; the
most common is the Center of Gravity method, in which the crisp
output value is calculated as the abscissa under the center of
gravity of the output fuzzy set (see Figure \ref{Mamdani}).
\end{enumerate}

In more complex technical applications of general function
approximation (like in complex control systems, signal and image
processing, etc.), two optional blocks are usually added to the
fuzzy inference engine \cite{Kosko,Kosko2,LeeC}:

0. Preprocessor, preceding the fuzzification module, performing
various kinds of normalization, scaling, filtering, averaging,
differentiation or integration of input data; and

4. Postprocessor, succeeding the defuzzification module,
performing the analog operations on output data.

Common fuzzy systems have a simple feedforward mathematical
structure, the so--called \textit{Standard Additive Model} (SAM,
for short), which aids the spread of applications. Almost all
applied fuzzy systems use some form of SAM, and some SAMs in turn
resemble the ANN models (see \cite{Kosko3}).

In particular, an \textit{additive fuzzy system}
$F:\Bbb{R}^{n}\rightarrow \Bbb{R}^{p}$ stores $m$ rules of the
patch form $A_{i}\times B_{i}\subset \Bbb{R}^{n}\times
\Bbb{R}^{p}$, or of the word form `\textbf{If} $X=A_{i}$
\textbf{Then} $Y=B_{i}$' and adds the `fired' Then--parts
$B_{i}^{^{\prime }}(x)$ to give the output set $B(x)$, calculated
as
\begin{equation}
B(x)=\sum_{i=1}^{n}w_{i}B_{i}^{^{\prime
}}(x)=\sum_{i=1}^{n}w_{i}\mu_{i}(x)B_{i}(x),\qquad i=1,...,n,
\label{adfuz}
\end{equation}
for a scalar rule weight $w_{i}>0$. The factored form
$B_{i}^{^{\prime }}(x)=\mu_{i}(x)B_{i}(x)$ makes the additive
system (\ref{adfuz})\ a SAM system. The fuzzy system $F$ computes
its output $F(x)$ by taking the centroid of the output set $B(x)$:
$F(x)=$ Centroid$(B(x))$. The \textit{SAM Theorem} then gives the
centroid as a simple ratio,
\[
F(x)=\sum_{i=1}^{n}p_{i}(x)c_{i},\qquad i=1,...,n,
\]
where the convex coefficients or discrete probability weights
$p_{i}(x)$ depend on the input $x$ through the ratios
\begin{equation}
p_{i}(x)=\frac{w_{i}\mu_{i}(x)V_{i}}{\sum_{k=1}^{n}w_{k}\mu_{k}(x)V_{k}},
\qquad i=1,...,n.  \label{pfuz}
\end{equation}
$V_{i}$ is the finite positive volume (or area if $p=1$ in the
codomain space $\Bbb{R}^{p}$) \cite{Kosko3},
\[
V_{i}=\int_{\Bbb{R}^{p}}b_{i}(y_{1},...,y_{p})dy_{1}...dy_{p}>0,
\]
and $c_{i}$ is the centroid of the Then--part set $B_{i}(x)$,
\[
c_{i}=\frac{\int_{\Bbb{R}^{p}}y\,b_{i}(y_{1},...,y_{p})dy_{1}...dy_{p}}{%
\int_{\Bbb{R}^{p}}b_{i}(y_{1},...,y_{p})dy_{1}...dy_{p}}.
\]

\subsection{Fuzzy Decision Making}

Recall that \textit{finite state machines} (FSMs) are simple
`machines' that have a finite number of states (or conditions) and
transition functions that determine how input to the system
changes it from one state to another \cite{NeuFuz}.

Fuzzy State Machines (FuSMs) are a modification of FSMs. In FuSMs,
the inputs to the system (that cause the transitions between
states) are not discrete. The real value of FuSMs comes from the
interaction of the system inputs. For example, a character in a
video game may of a simple combat scenario decides how aggressive
he will be depending on \textit{his health}, \textit{the enemy's
health}, and his \textit{distance from the enemy}. The combination
of these inputs cause the state transitions to happen. This can
result in very complex behaviors from a small set of rules. For example,

The \textit{health} variables have three sets:~ \textit{Near
death}, \textit{Good}, and \textit{Excellent}.

The \textit{distance} variable has three sets:~ \textit{Close},
\textit{Medium}, and \textit{Far}.

Finally, the output (\textit{aggressiveness}) has three sets:~
\textit{Run away}, \textit{Fight defensively}, and \textit{All out
attack!}.

\textit{Fuzzy Control Language} (FCL) is a standard for Fuzzy
Control Programming published by the International
Electrotechnical Commission (IEC).

\textit{Fuzzy--logic decision maker} (FLDM) breaks the decision
scenario down into small parts that we can focus on and input
easily. It then uses theoretically optimal methods of combining
the scenario pieces into a global interrelated whole with an
indication as to which alternative is the best within the
constraints and goals of the decision scenario.

The assumption in FLDM is that a judgment consists of a
\textit{known here and now} (the constraints), a \textit{hoped for
future there and then} (the goals), and \textit{various paths}
(the alternatives) for getting from the present here and now to
the future there and then. The problem is then the selection of
the path (alternative) that optimally supports the present
constraints and the future goals.

Decision making when faced with several alternatives, which
initially appear equally good or desirable, can be a time
consuming and often painful process. The FLDM overcomes the
(human) memory and processor limitations by allowing the decision
maker to selectively evaluate small amounts of the necessary
information at any one time (i.e., the fuzzy values of goal and
constraint satisfaction and simple, one-at-a-time paired
comparisons). Then, when it becomes necessary to evaluate all the
pertinent data, the computer can be utilized to perform the
decision task in a straight forward manner.

\subsection{Fuzzy Logic Control}
\label{fzcon}

The most common and straightforward applications of fuzzy logic
are in the domain of control \cite{Kosko,Kosko2,LeeC,Dote}. Fuzzy
control is a nonlinear control method based on fuzzy logic. Just
as fuzzy logic can be described simply as computing with words
rather than numbers, fuzzy control can be described simply as
control with sentences rather than differential equations.

A fuzzy controller is based on the fuzzy inference engine, which
acts either in the feedforward or in the feedback path, or as a
supervisor for the conventional PID controller.

A fuzzy controller can work either directly with fuzzified
dynamical variables, like direction, angle, speed, or with their
fuzzified errors and rates of change of errors. In the second case
we have rules of the form:
\begin{enumerate}
    \item If error is $Neg$ and change in error is $Neg$ then output is
$NB$.
    \item If error is $Neg$ and change in error is $Zero$ then output is $NM$.
\end{enumerate}

The collection of rules is called a rule base. The rules are in
$IF-THEN$
format, and formally the $IF-$side is called the condition and the $THEN-$%
side is called the conclusion (more often, perhaps, the pair is
called antecedent - consequent). The input value $Neg$ is a
linguistic term short
for the word Negative, the output value $NB$ stands for $Negative\_Big$ and $%
NM$ for $Negative\_Medium$. The computer is able to execute the
rules and compute a control signal depending on the measured
inputs error and change in error.

The rulebase can be also presented in a convenient form of one or
several rule matrices, the so--called $FAM-$matrices, where $FAM$
is a shortcut for Kosko's \textit{fuzzy associative memory}
\cite{Kosko,Kosko2} (see Figure 4). For example, a $9\times 9$
graded FAM matrix can be defined in a symmetrical weighted form:

{\
\begin{equation*}
\text{FAM}=\left(
\begin{tabular}{ccccc}
0.6S4 & 0.6S4 & 0.7S3 & ... & CE \\
0.6S4 & 0.7S3 & 0.7S3 & ... & 0.9B1 \\
0.7S3 & 0.7S3 & 0.8S2 & ... & 0.9B1 \\
... & ... & ... & ... & 0.6B4 \\
CE & 0.9B1 & 0.9B1 & ... & 0.6B4%
\end{tabular}%
\ \ \right),
\end{equation*}%
}\noindent in which the vector of nine linguistic variables
$L^{9}$ partitioning the \textit{universes of discourse} of all
three variables (with trapezoidal or Gaussian bell--shaped
\textit{membership functions}) has the form
\begin{equation*}
L^{9}=\{S4,S3,S2,S1,CE,B1,B2,B3,B4\}^{T},
\end{equation*}%
to be interpreted as: `small 4', ... , `small 1', `center', `big
1', ... , `big 4'. For example, the left upper entry $(1,1)$ of
the FAM matrix means: IF red is S4 and blue is S4, THEN result is
0.6S4; or, entry $(3,7)$ means: IF red is S2 and blue is B2, THEN
result is center, etc.

Here we give design examples for three fuzzy controllers.

\paragraph{Temperature Control System.}

In this simple example, the input linguistic variable is

$temperature\_error=desired\_temperature-current\_temperature$.
The two output linguistic variables are: $hot\_fan\_speed$, and
$cool\_fan\_speed$. The universes of discourse, consisting of
membership functions, i.e., overlapping triangular--trapezoidal
shaped intervals, for all three variables are:

$invar$:
$temperature\_error=\{Negative\_Big,Negative\_Medium,\newline
Negative\_Small,Zero,Positive\_Small,Positive\_Medium,Positive\_Big\}$,
with the range $[-110,110]$ degrees;

$outvars$: $hot\_fan\_speed$ and $cool\_fan\_speed=\{zero,low,medium,high,%
\newline
very\_high\}$, with the range $[0,100]$ rounds-per-meter.

\paragraph{Car Anti--Lock Braking System.}

The fuzzy---logic controller for the car anti--lock braking system
consists of the following \textit{input variables}:\bigbreak

\textit{slip\_r }(rear\_wheels\_slip),

\textit{slip\_fr} (front\_right\_wheel\_slip),

\textit{slip\_fl} (front\_left\_wheel\_slip),

\bigbreak\noindent with their membership functions:\bigbreak

\textit{NZ = Near\_Zero, \qquad  OP = Optimal, \qquad  AO =
Above\_Optimal,}

\bigbreak\noindent and the following \textit{output
variables}:\bigbreak

\textit{bp\_r }(rear\_wheels\_brake\_pressure),

\textit{bp\_fr} (front\_right\_brake\_pressure),

\textit{bp\_fl} (front\_left\_brake\_pressure),

\bigbreak\noindent with their membership functions:\bigbreak

\textit{LW = Low, \qquad  MD = Medium, \qquad  HG = High.}

\bigbreak\noindent The \textit{inference rule--base} for this
example consists of the following fuzzy implications:\bigbreak

IF \quad slip\_fl is NZ \quad THEN \quad bp\_fl is MD;

IF \quad slip\_fr is NZ \quad THEN \quad bp\_fr is MD;

IF \quad slip\_r is NZ \quad THEN \quad bp\_r is MD;

IF \quad slip\_fl is OP \quad THEN \quad bp\_fl is HG;

IF \quad slip\_fr is OP \quad THEN \quad bp\_fr is HG;

IF \quad slip\_r is OP \quad THEN \quad bp\_r is HG;

IF \quad slip\_fl is AO \quad THEN \quad bp\_fl is LW;

IF \quad slip\_fr is AO \quad THEN \quad bp\_fr is LW;

IF \quad slip\_r is AO \quad THEN \quad bp\_r is LW.\bigbreak

\paragraph{Truck Backer--Upper Steering Control System.}

\begin{figure}[tbh]
\centerline{\includegraphics[width=11cm]{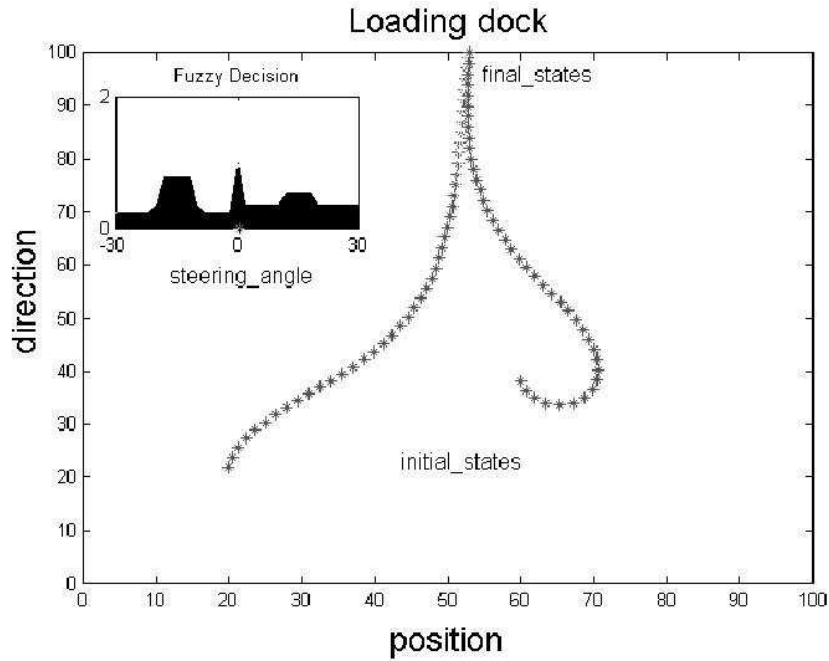}}
\caption{Truck backer--upper steering control system.}
\label{TruckBackUp}
\end{figure}

In this example there are two input linguistic variables: position
and direction of the truck, and one output linguistic variable:
steering angle (see Figure \ref{TruckBackUp}). The universes of
discourse, partitioned by overlapping triangular--trapezoidal
shaped intervals, are defined as:

$invars$: $position=\{NL,NS,ZR,PS,PL\}$, ~~and

$direction= \{NL,NM,NS,ZR,PS,PM,PL\}$, where $NL$ denotes Negative\_Large, $%
NM$ is Negative\_Medium, $NS$ is Negative\_Small, etc.

$outvar$: $steering\_angle=\{NL,NM,NS,ZR,PS,PM,PL\}$.

The rule--base is given as:
\begin{center}

IF direction is $NL$ and position is $NL$, THEN steering angle is
$NL$;

IF direction is $NL$ and position is $NS$, THEN steering angle is
$NL$;

IF direction is $NL$ and position is $ZR$, THEN steering angle is
$PL$;

IF direction is $NL$ and position is $PS$, THEN steering angle is
$PL$;

IF direction is $NL$ and position is $PL$, THEN steering angle is
$PL$;

IF direction is $NM$ and position is $NL$, THEN steering angle is
$ZR$;

. . . . . . . . . . . . .

IF direction is $PL$ and position is $PL$, THEN steering angle is
$PL$. \end{center}

\begin{figure}[tbh]
\centerline{\includegraphics[width=11cm]{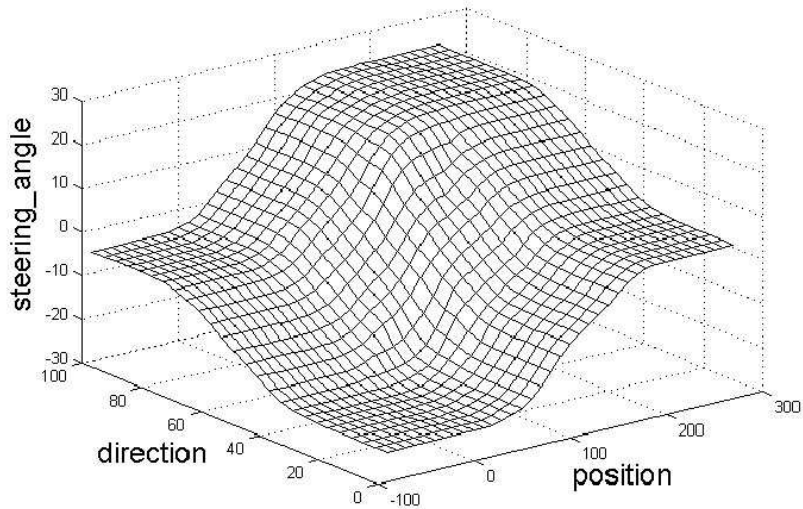}}
\caption{Control surface for the truck backer--upper steering
control system. } \label{CtrlSurf}
\end{figure}

The so--called \textit{control surface} for the truck
backer--upper steering control system is depicted in Figure
\ref{CtrlSurf}.

\subsubsection{Characteristics of Fixed Fuzzy Control}

Fuzzy logic offers several unique features that make it a
particularly good choice for many control problems, among them
\cite{LeeC,Dote}:
\begin{enumerate}
    \item It is inherently robust since it does not require precise,
noise--free inputs and can be programmed to fail safely if a
feedback sensor quits or is destroyed. The output control is a
smooth control function despite a wide range of input variations.
    \item Since the fuzzy logic controller processes user--defined rules
governing the target control system, it can be modified and
tweaked easily to improve or drastically alter system performance.
New sensors can easily be incorporated into the system simply by
generating appropriate governing rules.
    \item Fuzzy logic is not limited to a few feedback inputs and one or two
control outputs, nor is it necessary to measure or compute
rate--of--change parameters in order for it to be implemented. Any
sensor data that provides some indication of a systems actions and
reactions is sufficient. This allows the sensors to be inexpensive
and imprecise thus keeping the overall system cost and complexity
low.
    \item Because of the rule-based operation, any reasonable number of
inputs can be processed (1--8 or more) and numerous outputs (1--4
or more) generated, although defining the rulebase quickly becomes
complex if too many inputs and outputs are chosen for a single
implementation since rules defining their interrelations must also
be defined. It would be better to break the control system into
smaller chunks and use several smaller fuzzy logic controllers
distributed on the system, each with more limited
responsibilities.
    \item Fuzzy logic can control nonlinear systems that would be difficult or
impossible to model mathematically. This opens doors for control
systems that would normally be deemed unfeasible for automation.
\end{enumerate}

A \textit{fuzzy logic controller} is usually designed using the
following steps:
\begin{enumerate}
    \item Define the control objectives and criteria: What am I trying to control?
What do I have to do to control the system? What kind of response
do I need? What are the possible (probable) system failure modes?
    \item Determine the input and output relationships and choose a minimum number
of variables for input to the fuzzy logic engine (typically error
and rate--of--change of error).
    \item Using the rule--based structure of fuzzy logic, break the
control problem down into a series of \emph{IF X AND Y THEN Z}
rules that define the desired system output response for given
system input conditions. The number and complexity of rules
depends on the number of input parameters that are to be processed
and the number fuzzy variables associated with each parameter. If
possible, use at least one variable and its time derivative.
Although it is possible to use a single, instantaneous error
parameter without knowing its rate of change, this cripples the
systems ability to minimize overshoot for a step inputs.
    \item Create fuzzy logic membership functions that define the meaning (values)
of Input/Output terms used in the rules.
    \item Test the system, evaluate the results, tune the rules and membership
functions, and re-test until satisfactory results are obtained.
\end{enumerate}

Therefore, fuzzy logic does not require precise inputs, is
inherently robust, and can process any reasonable number of inputs
but system complexity increases rapidly with more inputs and
outputs. Distributed processors would probably be easier to
implement. Simple, plain--language rules of the form \emph{IF X
AND Y THEN Z} are used to describe the desired system response in
terms of linguistic variables rather than mathematical formulas.
The number of these is dependent on the number of inputs, outputs,
and the designers control response goals. Obviously, for very
complex systems, the rule--base can be enormous and this is
actually the only drawback in applying fuzzy logic.

\subsubsection{Pro and Contra Fuzzy Logic Control}

According to \cite{Abramovitch} there are the following pro and
contra arguments regarding fuzzy logic control:
\begin{enumerate}
    \item Fuzzy logic control is more useful than its detractors claim.
    \item Fuzzy logic control is less useful than its proponents claim.
    \item Fuzzy logic does not generate a control law. It maps an
existing control law from one set of rules into a logic set.
    \item Fuzzy logic control is most useful in `common sense' control
situations, i.e., ones where it might be difficult to write down
the equations of motion, but a human would know how to control it.
Examples of this are the `truck backer upper', car parking, train
control, and helicopter control problems.
    \item Fuzzy logic sets effectively quantize their input and output
space. However, the quantization intervals are rarely uniform.
    \item In most fuzzy logic control success stories the sample rates
are incredibly high relative to the dynamics of the system. Much
of their success is because of this.
\end{enumerate}

Most of the examples of fuzzy logic control being successfully
applied fall into the category of things that humans do well
\cite{Abramovitch,Cox92,Schwartz}.

Recall that in Japan, there is a train (Sendai subway), which is
controlled by fuzzy logic. The train pulls into the station within
a few inches of its target. More accurate, but nevertheless
replacing human control \cite{Cox92}.

Also in Japan, there are experiments in controlling a small model
helicopter (Spectrum, July 1992) via radio control. The helicopter
can respond to commands such as take off and land, hover, forward,
backwards, left and right \cite{Abramovitch,Cox92,Schwartz}.

Proponents assert that a conventional control scheme would be
incredibly hard to design because it would be really tough to
model the helicopter dynamics. The `model free' nature of fuzzy
logic control makes the problem trivial. This might be true, at
least from a practical application point of view, but it obscures
some key facts \cite{Abramovitch}:
\begin{enumerate}
    \item The model helicopter was designed so that a human operator with
a joystick could control it, i.e., it was designed to respond well
to intuitive control rules. Because of this, the helicopter has
been designed to be very robust to imprecision. (Robustness to
imprecision is one of the features that many proponents claim
fuzzy logic brings to the problem. It is possible that this
feature is more a feature of the dynamic system than of fuzzy
logic itself. In fact, L. Zadeh, the creator of fuzzy logic,
points out that fuzzy logic takes advantage of a system's inherent
robustness to imprecision rather than creating a robustness to
imprecision).
    \item The human operator has an implicit model in his mind of the
input-output behavior of the helicopter. This is how he generates
his control law for using the joystick.
    \item Fuzzy logic maps the human's control law and therefore is based on the
human's implicit model of the helicopter. This in turn works
because the helicopter was designed to be robust to human control
actions.
    \item The human being's `bandwidth' is quite low, certainly less
than 100 Hz. Furthermore, it is unlikely that a toy helicopter, a
train, or a truck would respond to anything about 1 Hz and
certainly not 10 Hz. (Since it must be an issue in every digital
control problem and since any implementation of fuzzy logic
control involves using some digital processor, the natural
conclusion is that the sample rates are chosen so high above the
system time constants that they seemingly stop being important.)
\end{enumerate}

The train control problem, as well as the car parking and truck
backer upper problem are all described by (1-4) above. So we can
conclude that high sample rates are an inherent part of using
fuzzy logic. The seemingly unimportant high sample rate may be
precisely why the simple control rules work well. Fast sampling
does lead to a greater computational burden. However, the
computational cost many be offset by being able to use a simpler
control law.

If we look in any fuzzy logic article we will see a picture of
membership functions for fuzzy sets (see, e.g., \cite{Cox92}).
These sets effectively quantize the interval that they are on:
they span the space so that any value on the line must fall into
at least one of the sets. However, they do not behave quite like
what we think of as quantizers since a particular value can be a
member of more than one set. The sets are typically fairly coarse
in terms of what we would consider effective quantization.
Combinations of these coarse quantizers provide various fuzzy
conditions. The coarse quantizations and simple rules may offset
the higher sample rate requirement.

In summary, fuzzy logic does not generate a control law, it merely
maps a law from one form to another. The simple rules for train
control or truck backing up are not generated by fuzzy logic
control. These are already present in the mind of the human
operator. Fuzzy logic merely maps the intuitive rules into a
computer program.

What seems to be the newest feature of fuzzy logic control is that
because the borders are fuzzy, more than one logic state can be
true to some degree. This allows for a smooth transition between
one control action and another, since they can both go on but at
different activation levels, or gain. Quite often control systems
have different operating regimes. Handling the transitions between
these tends to be ad hoc. Things, which are already ad hoc, are
perfect candidates for using fuzzy logic. Thus, fuzzy logic might
be a good solution for smoothly switching a control system from
one operating regime to another. In the transition, both control
laws would be active, but their outputs would be scaled by the how
much the system is in one regime or another. Clearly, this means
that both control laws would have to be run in parallel during the
transition.

On the other hand, quite a lot has been said about the
\textit{model--free} nature of a fuzzy logic control system. The
notion is that rather than trying to construct these complicated
dynamic models for a system, the `simple fuzzy rules' allow the
designer to design a control system. Clearly, this hides the
notion that buried in those `simple fuzzy rules' is an implicit
model of the system. Following \cite{Abramovitch}, we believe that
\textit{no intelligent action is possible without a model}. Any
general behavior trend constitutes a model, whether explicit (e.g.
dynamic systems model) or implicit (i.e. as encompassed in the
fuzzy logic rules).

Another general idea that seems to permeate the fuzzy logic
control hype is the notion that someone with very little skill can
design a controller using fuzzy logic, while using classical
control takes years of training.

In fact, the advantages and disadvantage of fuzzy systems result
of the fact that fuzzy logic represents a decision making process.
In control field, this provides a wide range of viable ways to
solve naturally control problems while the basic control knowledge
is not needed \cite{Vibet}.

Another thing to point out is that usually the fuzzy logic rules
use more external sensors, including acceleration, velocity, and
the position information. So they naturally perform better than
conventional controllers (position feedback loop) based only on
position sensors.

Many proponents of fuzzy logic control argue that fuzzy logic
works much better than conventional control when the system is
nonlinear. However, the conventional controller they are comparing
it to is a PID controller based on a linear system model.

In the sense that the fuzzy logic rules encompass a better model
(implicit but there) of the system than an inappropriately applied
linear model, the fuzzy logic rules will work better. Recall that
the linear model has its faults as well. If a control system is
designed using a linear model that doesn't characterize the system
behavior well, then the control system will probably fail to work
well. However, a fair comparison would be one made between a fuzzy
logic controller and a nonlinear state feedback controller that
measures all the same variables at the same sampling rate as the
fuzzy logic controller. If such a comparison is made there is no
guarantee that the fuzzy logic controller will work better.\bigbreak\bigbreak

\section{Adaptive Fuzzy Control in Human Operator Modeling}

\subsection{Neuro--Fuzzy Hybrid Systems}

In many applications, desired system behavior is partially
represented by data sets. In control systems, these data sets may
represent operational states. In decision support systems and data
analysis applications, these data sets may represent sample cases.

Discussing the respective strengths and weaknesses of fuzzy logic
and neural net technology, a simple comparison indicates that the
strongest benefit of a neural net is that it can automatically
learn from sample data. However, a neural net remains a black box,
thus manual modification and verification of a trained net is not
possible in a direct way.

This is where fuzzy logic excels: In a fuzzy logic system, any
component is defined as close as possible to human intuition,
making it very easy to manually modify and verify a designed
system. However, fuzzy logic systems can not automatically learn
from sample data.

This is where neuro--fuzzy system provides `the best of both
worlds'. Take the explicit representation of knowledge in
linguistic variables and rules from fuzzy logic and add the
learning approach used with neural nets. In the neuro--fuzzy
system, both fuzzy rules and membership functions are adjusted by
some form of backpropagation learning to adapt the system behavior
according to the sample data.

The neuro--fuzzy system can also be used to optimize existing
fuzzy logic systems. Starting with an existing fuzzy logic system,
the neuro--fuzzy system interactively tunes rule weights and
membership function definitions so that the system converges to
the behavior represented by the data sets.

To distinguish between more and less important rules in the
knowledge base, we can put weights on them. Such weighted
knowledge base can be then trained by means of artificial neural
networks. In this way we get \emph{hybrid neuro--fuzzy trainable
expert systems.}

\begin{figure}[tbh]
\centerline{\includegraphics[width=11cm]{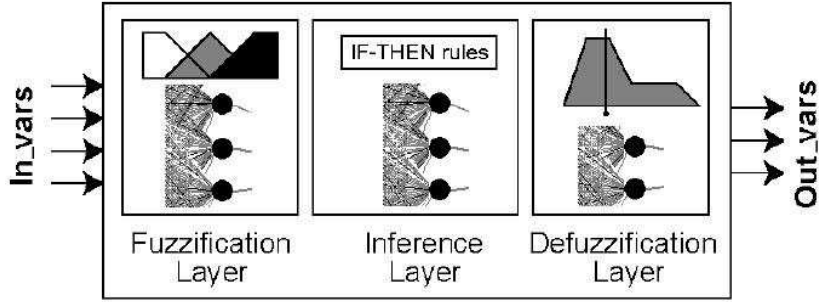}}
\caption{Neuro--fuzzy inference engine.}
\label{NeuroFuzzyInferEng}
\end{figure}

Another way of the hybrid neuro--fuzzy design is the fuzzy
inference engine such that each module is performed by a layer of
hidden artificial neurons, and ANN--learning capability is
provided to enhance the system knowledge (see Figure
\ref{NeuroFuzzyInferEng}).

Again, the fuzzy control of the backpropagation learning can be
implemented as a set of heuristics in the form of fuzzy $IF-THEN$
rules, for the purpose of achieving a faster rate of convergence.
The heuristics are driven by the behavior of the instantaneous sum
of squared errors.

As another alternative, we can consider the well--known fuzzy
ARTMAP system, which is essentially a clustering algorithm (vector
quantizer), with supervision that redirects training inputs which
would be grouped in an incorrect category to a different cluster.
A fuzzy ARTMAP system consists of two fuzzy ART modules, each of
which clusters vectors in an unsupervised fashion, linked by a map
field. Fuzzy ART clusters vectors based on two separate distance
criteria, \textit{match} and \textit{choice}. For more details,
see \cite{fuzzyART}.

Finally, most \textit{feedback fuzzy systems} are either discrete
or continuous generalized SAMs \cite{Kosko3}, given respectively
by
\[
x(k+1)=\sum_{i=1}^{n}p_{i}(x(k))B_{i}(x(k)),\qquad \text{or}\qquad
\dot{x}(t)=\sum_{i=1}^{n}p_{i}(x(t))B_{i}(x(t)),
\]
with coefficients $p_{i}$\ given by (\ref{pfuz}) above.

\subsection{Neuro--Fuzzy--Fractal Operator Control}

Although the general concept of learning, according to the
schematic recursion
$$NEW\,\,VALUE_{t_{n+1}}\,=\,OLD\,\,VALUE_{t_n}\,+\,INNOVATION
$$
-- can be implemented in the framework of nonlinear control theory
(as seen in the previous subsection), its natural framework is
artificial intelligence.

For the purpose of neuro--fuzzy--fractal control \cite{NeuFuz,ComputMind}, the general
model for a nonlinear plant can be modified as
\cite{nff1,nff2}
\begin{eqnarray}
\dot{x} &=&f_{1}(x,D,\alpha )-\beta f_{2}(x,D,\alpha ),  \label{nf1} \\
\dot{y} &=&\beta f_{2}(x,D,\alpha ),  \notag
\end{eqnarray}%
where $x\in \mathbb{R}^{n}$ is a vector of state variables, $y\in \mathbb{R}%
^{m}$ is a vector of the system outputs, $\beta \in \mathbb{R}$ is
a constant measuring the efficiency of the conversion process,
$D\in (0,3)$ is the \textit{fractal dimension} of the process, and
$\alpha \in \mathbb{R}$ is a fuzzy--inference selection parameter.

For a complex dynamical system it may be necessary to consider a
set of mathematical models to represent adequately all of possible
dynamic behaviors of the system. In this case, we need a decision
scheme to select the appropriate model to use according to the
linguistic value of a selection parameter. We use a \textit{fuzzy
inference system} for differential equations to achieve the model
selection. We have fuzzy rules of the form:\footnote{%
For programming purposes, recall that basic logical control
structures in the pseudocode include IF--THEN and SWITCH
statements, respectively defined as:

IF--THEN
\par
if ((condition1) \TEXTsymbol{\vert}\TEXTsymbol{\vert}
(condition2))
\par
\qquad \{action1;
\par
\} else if ((condition3) \&\& (condition4)) \{
\par
\qquad action2;
\par
\} else \{
\par
\qquad default action;
\par
\}
\par
\bigskip
\par
SWITCH
\par
switch (condition) \{
\par
case 1:
\par
\qquad action1;
\par
\qquad break;
\par
case 2:
\par
\qquad action2;
\par
\qquad break;
\par
default:
\par
\qquad default action;
\par
\qquad break;
\par
\}}

\begin{center}
IF $\alpha $\ is $A_{1}$ AND $D$ is $B_{1}$ THEN $M_{1}$

...\qquad \qquad \qquad ...\qquad \qquad \qquad ...

IF $\alpha $\ is $A_{n}$ AND $D$ is $B_{n}$ THEN $M_{n}$
\end{center}

\noindent where $A_{1},...,A_{n}$ are linguistic values for
$\alpha
,B_{1},...,B_{n}$ are linguistic values for the fractal dimension $D$, and $%
M_{1},...,M_{n}$ are mathematical models of the form given by
\ref{nf1}. The selection parameter $\alpha $ represents the
environment variable, like temperature, humidity, etc.

Following \cite{nff1,nff2}, we combine adaptive model--based
control using neural networks with the method for model selection
using fuzzy logic and fractal theory, to obtain a hybrid
neuro--fuzzy--fractal method for control of nonlinear plants. This
general method combines the advantages of neural networks (ability
for identification and control) with the advantages of fuzzy logic
(ability for decision and use of expert knowledge) to achieve the
goal of robust adaptive control of nonlinear dynamic plants. We
also use the fractal dimension to characterize the plant--output
processes in modeling these dynamical systems.

\subsubsection{Fractal Dimension for Machine Output Identification}

The experimental identification of a nonlinear biologic transducer
is often approached via consideration of its response to a
stochastic test ensemble, such as \textit{Gaussian white noise}
\cite{Marmarelis}. In this approach, the input--output
relationship a deterministic transducer is described by an
orthogonal series of functionals. Laboratory implementation of
such procedures requires the use of a particular test signal drawn
from the idealized stochastic ensemble; the statistics of the
particular test signal necessarily deviate from the statistics of
the ensemble. The notion of a \textit{fractal dimension}
(specifically the capacity dimension) is a means to characterize a
complex time series. It characterizes one aspect of the difference
between a specific example of a test signal and the test ensemble
from which it is drawn: the fractal dimension of ideal Gaussian
white noise is infinite, while the fractal dimension of a
particular test signal is finite. The fractal dimension of a test
signal is a key descriptor of its departure from ideality: the
fractal dimension of the test signal bounds the number of terms
that can reliably be identified in the orthogonal functional
series of an unknown transducer \cite{Victor}.

\paragraph{Definition of the fractal dimension.}

Recall that for a smooth (i.e., nonfractal) line, an approximate length $%
L(r) $ is given by the minimum number $N$ of segments of length
$r$ needed to cover the line, $L(r)=Nr$. As $r$ goes to zero,
$L(r)$ approaches a finite limit, the length $L$ of the curve.
Similarly one can define the area $A$ or the volume $V$ of
nonfractal objects as the limit of an integer power law of $r$,
\begin{equation*}
A=\lim_{r\rightarrow 0}Nr,\text{ \ \ \ \ \ \ \ \
}V=\lim_{r\rightarrow 0}Nr^{3},
\end{equation*}%
where the integer exponent is the Euclidean dimension $E$ of the
object.

This definition can not be used for fractal objects: as $r$ tends
to $0$, we enter finer and finer details of the fractal and the
product $Nr^{E}$ may
diverge to infinity. However, a real number $D$ exists so that the limit of $%
Nr^{D}$ stays finite. This exponent is called \textit{Hausdorff dimension} $%
D_{H}$, defined by%
\begin{equation*}
D_{H}=\lim_{r\rightarrow 0}\frac{\log N}{\log (1/r)}.
\end{equation*}%
Another popular definition of dimension proposed for fractal
objects is the \textit{correlation dimension} $D_{2}$, given by
\begin{equation*}
D_{2}=\lim_{r\rightarrow 0}\frac{\log C(r)}{\log (r)},
\end{equation*}%
where $C(r)$ is the number of points which have a smaller
(Euclidean) distance than a given distance $r$. This measure is
widely used because it is easy to evaluate for experimental data,
when the fractal comes from a `dust' of isolated points. A method
for measuring $D_{2}$ of strange attractors can be found in
\cite{Grassberger}. $D_{2}$ may also be used to determine whether
a time--series derives from a random process or from a
deterministic chaotic system. $m-$dimensional data vectors are
constructed from $m$ measurements spaced equidistant in time, and
$D_{2}$ is evaluated for this $m-$dimensional set of points. If
the time--series is a random process, $D_{2}$ increases with $m$;
if the time--series is a deterministic signal, $D_{2}$ does not
increase further when the embedding dimension m exceeds $D_{2}$.
Thus a plot of the correlation dimension as a function of the
embedding dimension may easily show whether a signal is random
noise of deterministic chaos. Note that $D_{2}\leq D_{H}.$

\paragraph{Fractal behavior and singularities in time series.}

The functions $y(t)$ typically studied in mathematical analysis
are continuous and have continuous derivatives. Hence, they can be
approximated
in the vicinity of some time $t_{i}$ by Taylor series (or power series)%
\begin{equation}
y(t)=a_{0}+a_{1}(t-t_{i})+a_{2}(t-t_{i})^{2}+a_{3}(t-t_{i})^{3}+...
\label{f1}
\end{equation}%
For small regions around $t_{i}$, just a few terms of the expansion (\ref{f1}%
) are necessary to approximate the function $y(t)$. In contrast,
most time series $y(t)$ found in `real--life' applications appear
quite noisy). Therefore, at almost every point in time, they
cannot be approximated either by Taylor series (or by Fourier
series) of just a few
terms. Moreover, many experimental or empirical time series have \textit{%
fractal features}, i.e., for some times $t_{i}$, the series $y(t)$
displays \textit{singular} behavior \cite{Amaral1,Amaral1}. By
this, we mean that at those times ti, the signal has components
with non--integer powers of time
which appear as step-like or cusp-like features, the so--called \textit{%
singularities}, in the signal.

Formally, one can write
\begin{equation}
y(t)=a_{0}+a_{1}(t-t_{i})+a_{2}(t-t_{i})^{2}+a_{3}(t-t_{i})^{3}+...+a_{h}(t-t_{i})^{h_{i}}
\label{f2}
\end{equation}%
where $t$ is inside a small vicinity of $t_{i}$, and $h_{i}$ is a
non--integer number quantifying the local singularity of $y(t)$ at
$t=t_{i}$.

The next problem is to quantify the `frequency' in the signal of a
particular value $h$ of the singularity exponents $h_{i}$.
Different possibilities can be considered. For example, the set of
times with singular behavior \{$t_{i}$\} may be a finite fraction
of the time series and homogeneously distributed over the signal.
But \{$t_{i}$\} may also be an asymptotically infinitesimal
fraction of the entire signal and have a very heterogeneous
structure. That is, the set \{$t_{i}$\} may be a fractal itself.
In either case, it is useful to quantify the properties of the
sets
of singularities in the signal by calculating their fractal dimensions $%
D_{2} $ or $D_{H}$.

\paragraph{Fractal dimension of a machine output signal.}

This method uses the fractal dimension to make a unique
classification of the different types of machine behavior, because
different types of signals have different geometrical forms. The
problem is then of finding a one--to--one map between the
different types of machine behaviors and their corresponding
fractal dimension. The first step in obtaining this map is to find
experimentally the different geometrical forms for machine output
signals. The second step is to calculate the corresponding fractal
dimensions for these signals. This fractal dimension can be
calculated for a selected type of signals with several samples, to
obtain as a result a statistical estimation of the fractal
dimension and the corresponding error of the estimation. In order
to make an efficient use of this map between the different types
of machine behaviors and their corresponding estimated dimensions,
we need to implement it as a module in the computer program.

\subsubsection{Fuzzy Logic Model Selection for Dynamical Systems}

For a real--world dynamical system it may be necessary to consider
a set of mathematical models to represent adequately all of the
possible dynamic behaviors of the system. In this case, we need a
fuzzy decision procedure to select the appropriate model to use
according to the value of a selection parameter vector $\alpha $.
To implement this decision procedure, we need a fuzzy inference
system that can use differential equations as consequents. For
this purpose, we can follow the \textit{fuzzy decision system}
developed in \cite{nff1,nff2}, that can be considered as a
generalization of the classical Sugeno's inference system
\cite{Sug1,Sug2,Kosko}, in which differential equations as
consequents of the fuzzy rules are used instead of simple
polynomials like in the original Sugeno's method. Using this
method, a fuzzy model for a general dynamical system can be
expressed as follows \cite{NeuFuz}:

\begin{center}
IF $\alpha _{1}$\ is $A_{11}$ AND $\alpha _{2}$\ is $A_{12}$ ...
AND $\alpha _{m}$\ is $A_{1m}$ THEN $\dot{y}=f_{1}(y,\alpha )$

IF $\alpha _{1}$\ is $A_{21}$ AND $\alpha _{2}$\ is $A_{22}$ ...
AND $\alpha _{m}$\ is $A_{2m}$ THEN $\dot{y}=f_{2}(y,\alpha )$

...\qquad \qquad \qquad ...\qquad \qquad \qquad ...

IF $\alpha _{1}$\ is $A_{n1}$ AND $\alpha _{2}$\ is $A_{n2}$ ...
AND $\alpha _{m}$\ is $A_{nm}$ THEN $\dot{y}=f_{n}(y,\alpha )$
\end{center}

\noindent where $A_{ij}$ is the linguistic value of $\alpha _{j}$
for the $i$th rule, $\alpha =[\alpha _{1},...,\alpha _{m}]\in \mathbb{R}^{m}$, and $%
y\in \mathbb{R}^{p}$ is the output obtained by the numerical
solution of the corresponding differential equation (it is assumed
that each differential equation in this fuzzy model locally
approximates the real dynamical system over a neighborhood (or
region) of $\mathbb{R}^{m}$).

For example, if a complex dynamical system is modelled by using
four
different mathematical models $(M_{1},M_{2},M_{3},M_{4})$ of the form (\ref%
{nf1}), the decision scheme can be expressed as a single--input
fuzzy model \cite{nff1,nff2}

\begin{center}
IF $\alpha $\ is small THEN $\dot{y}=f_{1}(y,\alpha )$,

IF $\alpha$\ is regular THEN $\dot{y}=f_{2}(y,\alpha )$,

IF $\alpha $\ is medium THEN $\dot{y}=f_{3}(y,\alpha )$,

IF $\alpha $\ is large THEN $\dot{y}=f_{4}(y,\alpha )$.
\end{center}

\noindent where the output $y$ is obtained by the numerical
solution of the corresponding differential equation.

\subsubsection{Parametric Adaptive Control Using Neural Networks}

A feedforward neural network model takes an input vector $X$ and
produces an output vector $Y$. The input--output map
$NN:X\rightarrow Y$ is determined by the network architecture
(see, e.g., \cite{Kosko, Haykin}). The feedforward network
generally consists of at least three layers: one input layer, one
output layer, and one or more hidden layers. In our case, the
input layer with $y+1$ processing elements, i.e., one for each
predictor variable plus a processing element for the bias. The
bias element always has an input of one, $X_{y+1}=1$. Each
processing element in the input layer sends signals
$X_{i}\,(i=1,...,y+1)$ to each of the $q$ processing elements in
the hidden layer. The $q$ processing elements in the hidden layer
(indexed by $j=1,...,q$) produce an `activation' $a_{j}=F\left(
\sum w_{ij}X_{i}\right) $ where $w_{ij}$ are the weights
associated with the connections between the $y+1$ processing
elements of the input layer and the $j$th processing element of
the hidden layer. Once again, processing element $q+1$ of the
hidden layer is a bias element and always has an activation of
one, i.e. $a_{q+1}=1$. Assuming that the processing element in the
output layer is linear, the network model will
be%
\begin{equation}
Y_{t}=\sum_{l=1}^{p+1}\pi _{l}x_{it}+\sum_{j=1}^{p+1}\theta
_{j}F\left( \sum_{j=1}^{p+1}w_{ij}X_{it}\right).  \label{nf7}
\end{equation}%
Here $\pi _{l}$ are the weights for the connections between the
input layer and the output layer, and $\theta _{j}$ are the
weights for the connections between the hidden layer and the
output layer. The main requirement to be satisfied by the
activation function $F(\cdot)$ is that it be nonlinear and
differentiable. Typical functions used are the sigmoid,
$F(x)=1/(1+\exp (-x)) $ and hyperbolic tangent, $F(x)=(\exp
(x)-\exp (-x))/(\exp (x)+\exp (-x))$.

Feedforward neural nets are trained by supervised training, the
most popular being some form of the \textit{backpropagation}
algorithm. As the name suggests, the error computed from the
output layer is backpropagated through the network, and the
weights are modified according to their contribution to the error
function. Essentially, backpropagation performs a \textit{local
gradient search}, and hence its implementation does not guarantee
reaching a global minimum. A number of heuristics are available to
partly address this problem, for practical purpose the best one is
the Levenberg--Marquardt algorithm. Instead of distinguishing
between the weights of the different layers as in (\ref{nf7}), we
refer to them generically as $w_{ij}$ in the following. After some
mathematical simplification the weight change $\Delta w_{ij}$
equation suggested by backpropagation can be expressed as (see,
e.g., \cite{Haykin,Kosko})
\begin{equation}
\Delta w_{ij}=-\eta (\partial E_{1}/\partial w_{ij})+\theta \Delta
w_{ij}, \label{nf10}
\end{equation}%
where $\eta $ is the learning coefficient and $\theta $\ is the
momentum term. One heuristic that is used to prevent the neural
network from getting stuck at a local minimum is the random
presentation of the training data. Another heuristic that can
speed up convergence is the cumulative update of weights, i.e.,
weights are not updated after the presentation of each
input--output pair, but are accumulated until a certain number of
presentations are made, this number referred to as an `epoch'. In
the absence of the second term in (\ref{nf10}), setting a low
learning coefficient results in slow learning, whereas a high
learning coefficient can produce divergent behavior. The second
term in (\ref{nf10}) reinforces general trends, whereas
oscillatory behavior is cancelled out, thus allowing a low
learning coefficient but faster learning. Last, it is suggested
that starting the training with a large learning coefficient and
letting its value decay as training progresses speeds up
convergence.

Now, parametric adaptive control is the problem of controlling the
output of a system with a known structure but unknown parameters.
These parameters can be considered as the elements of a vector
$y$. If $y$ is known, the parameter vector of a controller can be
chosen as $\theta ^{\ast }$ so that the plant together with the
fixed controller behaves like a reference model described by a
difference (or differential) equation with constant coefficients
\cite{soft}. If $y$ is unknown, the vector $\theta (t)$ has to be
adjusted on--line using all the available information concerning
the system.

Two distinct approaches to the adaptive control of an unknown
plant are (i) direct control and (ii) indirect control. In direct
control, the parameters of the controller are directly adjusted to
reduce some norm of the output
error. In indirect control, the parameters of the plant are estimated as $%
y(t)$ at any time instant and the parameter vector $\theta (t)$ of
the controller is chosen assuming that $y(t)$ represents the true
value of the plant parameter vector. Even when the plant is
assumed to be linear and time--invariant, both direct and indirect
adaptive control results in non--linear systems.

When indirect control is used to control a nonlinear system, the
plant is parameterized using a mathematical model of the general
form (\ref{nf1}) and the parameters of the model are updated using
the identification error. The controller's parameters in turn are
adjusted by backpropagating the error (between the identified
model and the reference model outputs) through the identified
model.

\input{ivdiag1}
\section{Application: Fuzzy-Control Based Tennis Simulator}

In this section we present a fuzzy--logic model for the
\emph{tennis game}, consisting of two stages:
\emph{attack} (AT) and \emph{counter--attack} (CA). For technical details, see \cite{NeuFuz}.

\subsubsection{Attack Model: Tennis Serve}

\noindent{\bf A. Simple Attack: Serve Only.} The simple
AT--dynamics is represented by a single fuzzy associative memory
(FAM) map
$$
\underset{CAT}{TARGET}\quad\underset{FAM}{\cone{{\cal
F}^{AT}}}\quad\underset{CAT}{ATTACK}
$$
In the case of simple tennis serve, this AT--scenario reads
$$
\underset{OPPONENT-IN}{O\ni o_{m}}\quad\cone{{\cal
F}^{AT}}\quad\underset{SERVE-OUT}{SR\ni sr_{n}}
$$
where the two $n-$categories, $O_{\dim =2}\ni o_{m}$ and $SR_{\dim
=3}\ni sr_{n}$, contain the temporal fuzzy variables
$\{o_{m}=o_{m}(t)\}$ and $\{sr_{n}=sr_{n}(t)\}$, respectively
opponent--related (target information) and serve--related,
partitioned by overlapping Gaussians, $\mu(z)=\\exp \left[
\frac{-(z-m)^{2}}{2\sigma ^{2}}\right]$, and defined as: \bigbreak

$\underset{OPPONENT-IN}{O}:%
\begin{array}{l}
o_{1}=Opp.Posit.Left.Right:(center,medium,wide), \\
o_{2}=Opp.Antcp.Lft.Rght:(runCenter,stay,runWide), \hspace{1cm}\\
\end{array}%
$\bigbreak

$\underset{SERVE-OUT}{SR}:%
\begin{array}{l}
sr_{1}=1.Serve.Speed : (low, medium, high)\\
sr_{2}=2.Serve.Spin : (low, medium, high)\\
sr_{3}=3.Serve.Placement : (center, medium, wide)
\end{array}%
$\bigbreak In the fuzzy--matrix form this simple serve reads
{\small
$$
\overset{O:\,OPPONENT-IN}{\left[
\begin{array}{l}
o_{1}=Opp.Posit.Left.Right \\
o_{2}=Opp.Anticip.Left.Right
\end{array}%
\right] }\quad\cone{{\cal
F}^{AT}}\quad\overset{SR:\;SERVE-OUT}{\left[
\begin{array}{l}
sr_{1}=1.Serve.Speed \\
sr_{2}=2.Serve.Spin \\
sr_{3}=3.Serve.Place
\end{array}\right] }
$$}

\noindent{\bf B. Attack--Maneuver: Serve--Volley.} The generic
advanced AT--dynamics is given by a composition of FAM functors
$$
\underset{CAT}{TARGET}\quad\underset{FAM}{\cone{{\cal
F}^{AT}}}\quad\underset{CAT}{ATTACK}\quad
\underset{FAM}{\cone{{\cal G}^{AT}}}\quad \underset{CAT}{MANEUVER}
$$
In the case of advanced tennis serve, this AT--scenario reads
$$
\underset{OPPONENT-IN}{O\ni o_{m}}\cone{{\cal
F}^{AT}}\underset{SERVE-OUT}{SR\ni sr_{n}}{\cone{{\cal
G}^{AT}}}\quad \underset{RUN-VOLEY}{RV\ni rv_{p}}
$$
where the new $n-$category, $RV_{\dim =2}\ni rv_{p}$, contains the
opponent--anticipation driven volley--maneuver, expressed by fuzzy
variables $\{rv_{p}=rv_{p}(t)\}$, partitioned by overlapping
Gaussians and given by:\bigbreak

$\underset{RUN-VOLEY}{RV}:%
\begin{array}{l}
rv_{1}=RV.For : (baseLine, center, netClose) \\
rv_{2}=RV.L.R. : (left,center,right)
\end{array}%
$\bigbreak In the fuzzy--matrix form this advanced serve reads
{\small
$$
\overset{O:\,OPPONENT-IN}{\left[
\begin{array}{l}
o_{1}=Opp.Posit.L.R. \\
o_{2}=Opp.Anticip.L.R.
\end{array}%
\right] }\cone{{\cal F}^{AT}}\overset{SR:\;SERVE-OUT}{\left[
\begin{array}{l}
sr_{1}=1.Serve.Speed \\
sr_{2}=2.Serve.Spin \\
sr_{3}=3.Serve.Place
\end{array}\right] }\cone{{\cal G}^{AT}}\overset{RV:\;RUN-VOLEY}{\left[
\begin{array}{l}
rv_{1}=RV.For\\
rv_{2}=RV.L.R
\end{array}%
\right] }
$$
}

\subsubsection{Counter--Attack Model: Tennis Return}

\noindent{\bf A. Simple Return.} The simple CA--dynamics reads:
$$
\underset{CAT}{ATTACK}\quad \underset{FAM}{\cone{{\cal
F}^{CA}}}\quad \underset{CAT}{MANEUVER}\quad
\underset{FAM}{\cone{{\cal G}^{CA}}}\quad \underset{CAT}{RESPONSE}
\label{1}
$$
In the case of simple tennis return, this CA--scenario consists
purely of conditioned--reflex reaction, no decision process is
involved, so it reads:
$$
\underset{BALL-IN}{B\ni b_{{\cal K}}}\quad {\cone{{\cal
F}^{CA}}}\quad \underset{RUNNING}{R\ni r_{{\cal J}}}\quad
{\cone{{\cal G}^{CA}}}\quad \underset{SHOT-OUT}{S\ni s_{k}}
\label{2}
$$
where the $n-$categories $B_{\dim =5}\ni b_{{\cal K}},$ $R_{\dim =3}\ni r_{{\cal J}},$ $%
S_{\dim =4}\ni s_{k}$, contain the fuzzy variables $%
\{b_{{\cal K}}=b_{{\cal K}}(t)\},$ $\{r_{{\cal J}}=r_{{\cal
J}}(t)\}$ and $\{s_{k}=s_{k}(t)\}$, respectively defining the ball
inputs, our player's running maneuver and his shot--response,
{\cal K}.e., {\small
$$
\overset{B:\;BALL-IN}{\left[
\begin{array}{l}
b_{1}=Dist.L.R. \\
b_{2}=Dist.{\cal F}.B. \\
b_{3}=Dist.Vert \\
b_{4}=Speed \\
b_{5}=Spin%
\end{array}%
\right] }{\cone{{\cal F}^{CA}}}\overset{R:\;RUNNING}{\left[
\begin{array}{l}
r_{1}=Run.L.R. \\
r_{2}=Run.{\cal F}.B. \\
r_{3}=Run.Vert%
\end{array}%
\right] }{\cone{{\cal G}^{CA}}}\overset{S:\;SHOT-OUT}{\left[
\begin{array}{l}
s_{1}=Backhand \\
s_{2}=Forehand \\
s_{3}=Voley \\
s_{4}=Smash%
\end{array}%
\right] }
$$}
Here, the existence of efficient weapons within the
$\underset{SHOT-OUT}{S}$ arsenal--space, namely $s_{k}(t):
s_{1}=Backhand,s_{2}=Forehand,s_{3}=Voley$ and $s_{4}=Smash$, is
assumed.

The universes of discourse for the fuzzy variables $\{b_{{\cal K}}(t)\},$ $%
\{r_{{\cal J}}(t)\}$ and $\{s_{k}(t)\}$, partitioned by
overlapping Gaussians, are defined respectively as:\bigbreak

$\underset{BALL-IN}{B}:%
\begin{array}{l}
b_{1}=Dist.L.R.:(veryLeft,left,center,right,veryRight), \\
b_{2}=Dist.{\cal F}.B.:(baseLine,center,netClose), \\
b_{3}=Dist.Vert:(low,medium,high), \\
b_{4}=Speed:(low,medium,high), \\
b_{5}=Spin:(highTopSpin,lowTopSpin,flat,\\ \qquad\qquad\qquad
 lowBackSpin,highBackSpin).%
\end{array}%
$\bigbreak

$\underset{RUNNING}{R}:\\
\begin{array}{l}
r_{1}=Run.L.R.:(veryLeft,left,center,right,veryRight), \\
r_{2}=Run.{\cal F}.B.:(closeFront,front,center,back,farBack), \\
r_{3}=Run.Vert:(squat,normal,jump).%
\end{array}%
$\bigbreak

$\underset{SHOT-OUT}{S}:%
\begin{array}{l}
s_{1}=Backhand:(low,medium,high), \\
s_{2}=Forehand:(low,medium,high), \\
s_{3}=Voley:(backhand,block,forehand), \\
s_{4}=Smash:(low,medium,high).%
\end{array}%
$\bigbreak

\noindent{\bf B. Advanced Return.} The advanced CA--dynamics
includes both the information about the opponent and (either
conscious or subconscious) decision making. This generic
CA--scenario is formulated as the following composition + fusion
of FAM functors:
$$
\underset{CAT}{ATTACK}\underset{FAM}{\cone{{\cal F}^{CA}}}\underset{%
CAT}{MANEUV}\underset{FAM}{\cone{{\cal G}^{CA}}}\;\underset{CAT}{DECISION%
}\underset{FAM}{\cone{{\cal H}^{CA}}}\;\;\underset{CAT}{RESP}
\label{3} \putsquare <0`-1`0`0;10`500>(-1100,-550)[```;`{\cal
K}^{CA}`FAM`] \put(-1280,-600){$\underset{CAT}{TARGET}$}
$$
where we have added two new $n-$categories,
$\underset{CAT}{TARGET}$ and $\underset{CAT}{DECISION}$,
respectively containing information about the opponent as a
target, as well as our own aiming decision processes. In the case
of advanced tennis return, this reads:
$$
\underset{BALL-IN}{B\ni b_{{\cal K}}}\;\;{\cone{{\cal F}^{CA}}}\;\;%
\underset{RUNNING}{R\ni r_{{\cal J}}}\;\;{\cone{{\cal G}^{CA}}}\;\;%
\underset{DECISION}{D\ni d_{l}}\;\;{\cone{{\cal H}^{CA}}}\;\;%
\underset{SHOT-OUT}{S\ni s_{k}}  \label{4} \putsquare
<0`-1`0`0;10`500>(-1200,-550)[```;`{\cal K}^{CA}``]
\put(-1500,-580){$\underset{OPPONENT-IN}{O\ni o_{m}}$}
$$
where the two additional $n-$categories, $O_{\dim =4}\ni o_{m}$
and $D_{\dim =5}\ni d_{l}$, contain the fuzzy variables
$\{o_{m}=o_{m}(t)\}$ and $\{d_{l}=d_{l}(t)\}$, respectively
defining the opponent--related target information and the
aim--related decision processes, both partitioned by overlapping
Gaussians and defined as:\bigbreak

$\underset{OPPONENT-IN}{O}:%
\begin{array}{l}
o_{1}=Opp.Posit.L.R.:(left,center,right), \\
o_{2}=Opp.Posit.{\cal F}.B.:(netClose,center,baseLine), \\
o_{3}=Opp.Anticip.L.R.:(runLeft,stay,runRight), \\
o_{4}=Opp.Anticip.{\cal F}.B.:(runNet,stay,runBase).
\end{array}%
$\bigbreak

$\underset{DECISION}{D}:%
\begin{array}{l}
d_{1}=Aim.L.R.:(left,center,right), \\
d_{2}=Aim.{\cal F}.B.:(netClose,center,baseLine), \\
d_{3}=Aim.Vert:(low,medium,high), \\
d_{4}=Aim.Speed:(low,medium,high), \\
d_{5}=Aim.Spin:(highTopSpin,lowTopSpin,noSpin,\\
\qquad\qquad\qquad\qquad lowBackSpin,highBackSpin).%
\end{array}%
$\bigbreak

The corresponding fuzzy--matrices read: {\small
\begin{eqnarray*}
\overset{B:\;BALL-IN}{\left[
\begin{array}{l}
b_{1}=Dist.L.R. \\
b_{2}=Dist.{\cal F}.B. \\
b_{3}=Dist.Vert \\
b_{4}=Speed \\
b_{5}=Spin%
\end{array}%
\right] },\qquad \overset{R:\;RUNNING}{\left[
\begin{array}{l}
r_{1}=Run.L.R. \\
r_{2}=Run.{\cal F}.B. \\
r_{3}=Run.Vert%
\end{array}%
\right] },\qquad \overset{D:\;DECISION}{\left[
\begin{array}{l}
d_{1}=Aim.L.R. \\
d_{2}=Aim.{\cal F}.B. \\
d_{3}=Aim.Vert \\
d_{4}=Aim.Speed \\
d_{5}=Aim.Spin%
\end{array}%
\right] }, \\
{\left[ \overset{O:\,OPPONENT-IN}{
\begin{array}{l}
o_{1}=Opp.Posit.L.R. \\
o_{2}=Opp.Posit.{\cal F}.B. \\
o_{3}=Opp.Anticip.L.R. \\
o_{4}=Opp.Anticip.{\cal F}.B.%
\end{array}%
}\right]},\qquad \overset{S:\;SHOT-OUT}{ \left[
\begin{array}{l}
s_{1}=Backhand \\
s_{2}=Forehand \\
s_{3}=Voley \\
s_{4}=Smash%
\end{array}%
\right]. }
\end{eqnarray*}
}
\bigbreak\bigbreak

\section{Conclusion}

In this paper we have presented several control strategies for human operator and sport modelling. Roughly, they are deviled into fixed-fuzzy control methods and adaptive-fuzzy control methods. As an application of the presented fuzzy control approaches to sport modelling we have presented a fuzzy-control based tennis simulator, consisting of attack and counter-attack stages. This approach can be applied to all sport games.

\end{document}

%% file: ivdiag1.tex
% 0. META

\newcommand{\mcm}[3]{\newcommand{#1}[#2]{{\ensuremath{#3}}}}

% 1. PACKAGES

%\usepackage{latexsym}
%\usepackage{amssymb}

% 2. FORMATTING

% Characters

\mcm{\blank}{0}{(\emptybk)} \mcm{\dashbk}{0}{\mbox{---}}
\mcm{\emptybk}{0}{\:\:} \mcm{\hyph}{0}{\mbox{-}}

% Diagrams

\mcm{\diagspace}{0}{\mbox{\hspace{2em}}}

% Proofs

\newcommand{\done}{\hfill\ensuremath{\Box}}
%\newcommand{\pf}{\noindent\textbf{Proof}}

% References

\newcommand{\bref}[1]{(\ref{#1})}
\newcommand{\ucontents}[2]{\addcontentsline{toc}{#1}{\numberline{}{#2}}}

% 3. TYPOGRAPHIC

\mcm{\cat}{1}{\mc{#1}} \mcm{\fcat}{1}{\mb{#1}}
\mcm{\mc}{1}{\mathcal{#1}} \mcm{\mr}{1}{\mathrm{#1}}
\mcm{\mi}{1}{\mathit{#1}} \mcm{\mb}{1}{\mathbf{#1}}
\mcm{\scat}{1}{\Bbb{#1}} \mcm{\twid}{1}{\widetilde{#1}}

% 4. SINGLE SYMBOLS

% Set theory

\mcm{\elt}{0}{\in} \mcm{\sub}{0}{\,\subseteq\,}
\mcm{\such}{0}{\:|\:} \mcm{\without}{0}{\setminus}

% Category theory

\mcm{\atsr}{0}{\Box} \mcm{\eqv}{0}{\,\simeq\,}
\mcm{\iso}{0}{\,\cong\,}
\mcm{\of}{0}{\raisebox{0.2mm}{\ensuremath{\scriptstyle\circ}}}

% Misc

\mcm{\bdry}{0}{\partial}

% 5. SPECIAL EXPRESSIONS

% Funny letters

\mcm{\Bee}{0}{\cat{B}} \mcm{\Beep}{0}{\cat{B'}}
\mcm{\Eee}{0}{\cat{E}} \mcm{\Eeep}{0}{\cat{E'}}
\newcommand{\epsln}{\varepsilon}
\mcm{\Ess}{0}{\cat{S}} \mcm{\Tee}{0}{\cat{T}}
\mcm{\Teep}{0}{\cat{T'}} \mcm{\Stee}{0}{\scat{T}}
\mcm{\Steep}{0}{\scat{T'}}

% Stars and blobs

\mcm{\blbk}{0}{\blank^{\blob}}
\mcm{\blob}{0}{\scriptscriptstyle{\bullet}}
\mcm{\stbk}{0}{\blank^{*}} \mcm{\ubl}{0}{{}^{\blob}}
\mcm{\ust}{0}{{}^{*}}

% Monads

\mcm{\Cartpr}{0}{\pr{\Eee}{T}} \mcm{\Cartprp}{0}{\pr{\Eeep}{T'}}
\mcm{\Mnd}{0}{\triple{T}{\eta}{\mu}}
\mcm{\Zeropr}{0}{\pr{\Set}{\id}}

% Pasting diagrams

\mcm{\dopset}{0}{\ftrcat{\Delta^{\op}}{\Set}}
\mcm{\tropset}{0}{\ftrcat{\fcat{TR}^{\op}}{\Set}}

% 6. CONSTANTS

% Roman

\mcm{\cod}{0}{\mr{cod}} \mcm{\dom}{0}{\mr{dom}}
\mcm{\End}{0}{\mr{End}} \mcm{\Hom}{0}{\mr{Hom}}
\mcm{\ob}{0}{\mr{ob}\,} \mcm{\op}{0}{\mr{op}}

% Italic

\mcm{\comp}{0}{\mi{comp}} \mcm{\id}{0}{\mi{id}}
\mcm{\ids}{0}{\mi{ids}} \mcm{\mult}{0}{\mi{mult}}
%\mcm{\unit}{0}{\mi{unit}}

% Bold

\mcm{\Ab}{0}{\fcat{Ab}} \mcm{\Alg}{0}{\fcat{Alg}}
% \mcm{\Bicat}{0}{\fcat{Bicat}}
\mcm{\Bim}{1}{\fcat{Bim}(#1)} \mcm{\Cat}{0}{\fcat{Cat}}
\mcm{\Cay}{0}{\fcat{Cay}} \mcm{\Cpn}{1}{\pr{\Set/S_{#1}}{T_{#1}}}
\mcm{\fc}{0}{\fcat{fc}} \mcm{\fm}{0}{\fcat{fm}}
\mcm{\Graph}{0}{\fcat{Graph}} \mcm{\Gy}{0}{\fcat{Gy}}
\mcm{\Hpn}{1}{\pr{\Eee_{#1}}{P_{#1}}} \mcm{\Mon}{0}{\mb{Mon}}
\mcm{\Multicat}{0}{\fcat{Multicat}} \mcm{\One}{0}{\fcat{1}}
\mcm{\PD}{1}{\fcat{PD}_{#1}} \mcm{\Prof}{0}{\fcat{Prof}}
\mcm{\Set}{0}{\fcat{Set}} \mcm{\Span}{0}{\fcat{Span}}
\mcm{\Ssq}{0}{\fcat{Ssq}} \mcm{\Struc}{0}{\fcat{Struc}}
\mcm{\Sym}{0}{\fcat{Sym}} \mcm{\TR}{1}{\fcat{TR}(#1)}
\mcm{\Tr}{0}{\fcat{Tr}} \mcm{\Twocat}{0}{\fcat{2\hyph\Cat}}

% Blackboard bold

\mcm{\integers}{0}{\mathbb{Z}}

% 7. TUPLES

% Plain tuples

\mcm{\range}{2}{#1,\,\ldots\,,#2}
\mcm{\bftuple}{2}{\tuplebts{\range{#1}{#2}}}
\mcm{\tuple}{3}{\tuplebts{\range{#1,#2}{#3}}}
\mcm{\rttuple}{1}{\tuplebts{\,\ldots\,,#1}}
\mcm{\abftuple}{2}{\atuplebts{\range{#1}{#2}}}
\mcm{\atuple}{3}{\atuplebts{\range{#1,#2}{#3}}}
\mcm{\arttuple}{1}{\atuplebts{\,\ldots\,,#1}}
\mcm{\sqbftuple}{2}{\obt\range{#1}{#2}\cbt}
\mcm{\pr}{2}{\tuplebts{#1,#2}}
\mcm{\triple}{3}{\tuplebts{#1,#2,#3}}

% Homs

\mcm{\eend}{2}{#1[#2]} \mcm{\ehom}{3}{#1[#2,#3]}
\mcm{\ftrcat}{2}{[#1,#2]} \mcm{\homset}{3}{#1(#2,#3)}
\mcm{\multihom}{3}{#1(#2;#3)}
\mcm{\relhom}{5}{#1_{#2}(\range{#3}{#4};#5)}

% 8. ARROWS

% Single arrows

\mcm{\go}{0}{\rTo} \mcm{\goby}{1}{\rTo^{#1}}
\mcm{\goesto}{0}{\,\longmapsto\,} \mcm{\goiso}{0}{\goby{\diso}}
\mcm{\monic}{0}{\rMonic} \mcm{\og}{0}{\lTo}
\mcm{\ogby}{1}{\lTo^{#1}}

% Plural arrows

\mcm{\gph}{2}{\spn{#1}{T #2}{#2}} \mcm{\graph}{4}{\spaan{#1}{T
#2}{#2}{#3}{#4}} \mcm{\oppair}{2}{\stackrel{\rTo^{#1}}{\lTo_{#2}}}
\mcm{\parpair}{2}{\stackrel{\rTo^{#1}}{\rTo_{#2}}}
\mcm{\spn}{3}{#2 \og #1 \go #3} \mcm{\spaan}{5}{#2 \ogby{#4} #1
\goby{#5} #3}

% Slice objects

\mcm{\bktdvslob}{3}
    {\left(
    \begin{diagram}[height=1.5em]
    #1      \\
    \dTo>{\,#2} \\
    #3      \\
    \end{diagram}
    \right)}
\mcm{\slob}{3}{(#1 \goby{#2} #3)} \mcm{\vslob}{3}
    {\left.
    \begin{diagram}[height=1.5em]
    #1      \\
    \dTo>{\,#2} \\
    #3      \\
    \end{diagram}
    \right.}

% 9. PT DIAGRAMS

% Environments

\newenvironment{opetope}
    {\begin{diagram}[size=1em,abut,tight,noPS]}
    {\end{diagram}}
\newenvironment{slopeydiag}
    {\begin{diagram}[size=2em]}
    {\end{diagram}}
\newenvironment{tree}
    {\begin{diagram}[height=1em,width=.75em,abut,noPS,tight]}
    {\end{diagram}}
\newenvironment{triangdiag}
    {\begin{diagram}[width=1em,height=1.5em]}
    {\end{diagram}}

% Trees

\newcommand{\dn}{\dLine}
\mcm{\enode}{0}{\circ}
\newcommand{\lt}[1]{\ldLine(#1,2)}
\mcm{\nl}{1}{\stackrel{\textstyle #1}{\node}}
\mcm{\node}{0}{\bullet}
\newcommand{\rt}[1]{\rdLine(#1,2)}
\mcm{\utree}{0}{\node}

% Miscellaneous

\newcommand{\cnr}{} % for opetopes
\mcm{\diso}{0}{\sim}
\newcommand{\pullshape}
    {\setlength{\unitlength}{1em}
    \begin{picture}(2,5)(-1,-5)
    \put(0,-5){\line(1,1){1}}
    \put(0,-5){\line(-1,1){1}}
    \end{picture}}
\newcommand{\Spbk}{\overprint{\raisebox{-2.5em}{\pullshape}}}
\mcm{\vdiso}{0}{\wr}

% ********************************  E N D  ********************************

\newcommand{\fcm}{\vsn{FCM}}
\mcm{\nat}{0}{\mathbb{N}}
\newcommand{\absentpiccy}[1]{\mbox{\textsc{Picture: }\texttt{#1}}}
\newcommand{\piccy}[1]{\epsfig{file=#1}}

\mcm{\Onepr}{0}{\pr{\Graph}{\fc}}
\newlength{\nllwidth}
\newlength{\nllheight}
\newcommand{\stackbelow}[2]{%
\settowidth{\nllwidth}{\ensuremath{#1}\ensuremath{#2}}%
\settoheight{\nllheight}{\ensuremath{#2}}%
\addtolength{\nllheight}{.3ex}%
\mbox{%
\ensuremath{#1}%
\hspace{-.5\nllwidth}%
\raisebox{-1\nllheight}{\ensuremath{#2}}}}
\mcm{\nlal}{2}{\stackbelow{\nl{#1}}{#2}}
\mcm{\nll}{1}{\stackbelow{\node}{#1}} \mcm{\wun}{0}{\fcat{1}}
\mcm{\atuplebts}{1}{\langle #1 \rangle} \mcm{\tuplebts}{1}{(#1)}
\mcm{\bo}{0}{(} \mcm{\bc}{0}{)}
\mcm{\UBilax}{0}{\fcat{UBicat}_\mr{lax}}
\mcm{\UBiwk}{0}{\fcat{UBicat}_\mr{wk}}
\mcm{\UBistr}{0}{\fcat{UBicat}_\mr{str}}
\mcm{\Bilax}{0}{\fcat{Bicat}_\mr{lax}}
\mcm{\Biwk}{0}{\fcat{Bicat}_\mr{wk}}
\mcm{\Bistr}{0}{\fcat{Bicat}_\mr{str}} \mcm{\rotsub}{0}{\cup
\raisebox{0.1em}{$\scriptstyle{|}$}} \mcm{\pd}{0}{\fcat{pd}}
\mcm{\rep}{1}{\widehat{#1}} \mcm{\ovln}{1}{\overline{#1}}
\mcm{\Gph}{0}{\fcat{Gph}} \mcm{\tr}{0}{\fcat{tr}}
\newcommand{\UB}{\textbf{(UB)}}
\newcommand{\UF}{\textbf{(UF)}}
\newcommand{\CB}{\textbf{(CB)}}
\newcommand{\CF}{\textbf{(CF)}}
\mcm{\ladj}{0}{\,\dashv\,} \mcm{\zeropd}{0}{\node}
\newenvironment{ntdiag}
    {\begin{diagram}[size=1.5em,noPS]}  % added in noPS - looks nicer
                        % as diagonal arrow is
                        % shorter
    {\end{diagram}}
\mcm{\END}{0}{\fcat{End}} \mcm{\HOM}{0}{\fcat{Hom}}

%%%%%%%%%%%%%%%%%%%%%%%%%%%%%%%%%%%%%%%%%%%%%%%%%%%%%%%%%%%%%%%%%%%%%%%%%

% hdpics

% PRESENTATION COMMANDS

% New lengths

\newlength{\gwidth} % the width of a glob
\newlength{\gvert}  % the overall vertical measurement
\newlength{\gdrop}  % the distance a labelled glob protrudes below the
            % text-line
\newlength{\gbaredrop}  % the distance from the textline to the bottom of the
                % unadorned glob
\newlength{\goffset}    % the distance from the centre of the glob to the
                % textline
\newlength{\gtemp}  % temporary register

% The `present' commands

\newcommand{\present}[1]{%
\makebox[1\gwidth]{%
\rule[-1\gdrop]{0ex}{1\gvert}%
\raisebox{-1\gbaredrop}{#1}}}

\newcommand{\presentl}[1]{%
\makebox[1\gwidth][l]{%
\rule[-1\gdrop]{0ex}{1\gvert}%
\raisebox{-1\gbaredrop}{#1}}}

\newcommand{\presentr}[1]{%
\makebox[1\gwidth][r]{%
\rule[-1\gdrop]{0ex}{1\gvert}%
\raisebox{-1\gbaredrop}{#1}}}

% Dimension commands

\newcommand{\ginitdims}[2]{%        % GLOBULAR VERSION
\setlength{\unitlength}{1em}%       % unitlength = 1em
\setlength{\goffset}{.25\unitlength}%   % globular offset = .25em
\setlength{\gwidth}{#1\unitlength}% % width as specified
\setlength{\gvert}{#2\unitlength}%  % vert = #2
\setlength{\gdrop}{.5\gvert}%       %
\addtolength{\gdrop}{-1\goffset}%   %
\setlength{\gbaredrop}{1\gdrop}%    % gdrop = drop = half(vert) - offset
\addtolength{\gvert}{.6\unitlength}%    % total extra clearance of .6em...
\addtolength{\gdrop}{.3\unitlength}}    % ...half of which is at bottom

\newcommand{\cinitdims}[2]{%        % CELLULAR VERSION
\setlength{\unitlength}{1em}%       % unitlength = 1em
\setlength{\goffset}{.35\unitlength}%   % cellular offset = .35em
\setlength{\gwidth}{#1\unitlength}% % width as specified
\setlength{\gvert}{#2\unitlength}%  % vert = #2
\setlength{\gdrop}{.5\gvert}%       %
\addtolength{\gdrop}{-1\goffset}%   %
\setlength{\gbaredrop}{1\gdrop}%    % gdrop = drop = half(vert) - offset
\addtolength{\gvert}{.6\unitlength}%    % total extra clearance of .6em...
\addtolength{\gdrop}{.3\unitlength}}    % ...half of which is at bottom

\newcommand{\gsinitdims}[2]{%       % SMALL GLOBULAR VERSION
\setlength{\unitlength}{0.5em}%     % unitlength = 0.5em
\setlength{\goffset}{.25\unitlength}%   % globular offset = .25em
\setlength{\gwidth}{#1\unitlength}% % width as specified
\setlength{\gvert}{#2\unitlength}%  % vert = #2
\setlength{\gdrop}{.5\gvert}%       %
\addtolength{\gdrop}{-1\goffset}%   %
\setlength{\gbaredrop}{1\gdrop}%    % gdrop = drop = half(vert) - offset
\addtolength{\gvert}{.6\unitlength}%    % total extra clearance of .6em...
\addtolength{\gdrop}{.3\unitlength}}    % ...half of which is at bottom

\newcommand{\sidespic}[1]{%
\settowidth{\gtemp}{\ensuremath{#1}}%
\addtolength{\gwidth}{1\gtemp}}

\newcommand{\abovepic}[1]{%
\settoheight{\gtemp}{\ensuremath{#1}}%
\addtolength{\gvert}{1\gtemp}%
\settodepth{\gtemp}{\ensuremath{#1}}%
\addtolength{\gvert}{1\gtemp}}

\newcommand{\belowpic}[1]{%
\settoheight{\gtemp}{\ensuremath{#1}}%
\addtolength{\gvert}{1\gtemp}%
\addtolength{\gdrop}{1\gtemp}%
\settodepth{\gtemp}{\ensuremath{#1}}%
\addtolength{\gvert}{1\gtemp}%
\addtolength{\gdrop}{1\gtemp}}

% MISC COMMANDS

\newcommand{\cell}[4]{\put(#1,#2){\makebox(0,0)[#3]{\ensuremath{#4}}}}
\mcm{\zmark}{0}{\scriptstyle{\bullet}}

%%%%%%%%%%%%%%%%%%%%%%%%%%%%%%%%%%%%%%%%%%%%%%%%%%%%%%%%%%%%%%%%%%%%%%%%%%%
%
% GLOBULAR PICTURES
%

\newcommand{\pregfst}[1]{%
\begin{picture}(0.5,0.2)(-0.5,-0.2)%
% label
\cell{-0.1}{-0.2}{tr}{#1}%
% dot
\cell{0}{0}{c}{\zmark}%
\end{picture}}

\mcm{\gfst}{1}{%
\ginitdims{0.5}{0.4}%
\sidespic{#1}%
\belowpic{#1}%
\presentr{\pregfst{#1}}}

\newcommand{\preglst}[1]{%
\begin{picture}(0.5,0.2)(0,-0.2)%
% label
\cell{0.1}{-0.2}{tl}{#1}%
% dot
\cell{0.05}{0}{c}{\zmark}%
\end{picture}}

\mcm{\glst}{1}{%
\ginitdims{.5}{.4}%
\sidespic{#1}%
\belowpic{#1}%
\presentl{\preglst{#1}}}

\newcommand{\preglft}[1]{%
\begin{picture}(0,0.2)(0,-0.2)%
% label
\cell{-0.1}{-0.2}{tr}{#1}%
% dot
\cell{0.05}{0}{c}{\zmark}%
\end{picture}}

\mcm{\glft}{1}{%
\ginitdims{0}{.4}%
\belowpic{#1}%
\present{\preglft{#1}}}

\newcommand{\pregrgt}[1]{%
\begin{picture}(0,0.2)(0,-0.2)%
% label
\cell{0.1}{-0.2}{tl}{#1}%
% dot
\cell{0.05}{0}{c}{\zmark}%
\end{picture}}

\mcm{\grgt}{1}{%
\ginitdims{0}{.4}%
\belowpic{#1}%
\present{\pregrgt{#1}}}

\newcommand{\pregblw}[1]{%
\begin{picture}(0,0.3)(0,-0.3)
% label
\cell{0}{-0.3}{t}{#1}%
% dot
\cell{0.05}{0}{c}{\zmark}%
\end{picture}}

\mcm{\gblw}{1}{%
\ginitdims{0}{.6}%
\belowpic{#1}%
\present{\pregblw{#1}}}

\newcommand{\pregfbw}[1]{%
\begin{picture}(0,0.65)(0,-0.65)
% label
\cell{0}{-0.65}{t}{#1}%
% dot
\cell{0.05}{0}{c}{\zmark}%
\end{picture}}

\mcm{\gfbw}{1}{%
\ginitdims{0}{1.3}%
\belowpic{#1}%
\present{\pregfbw{#1}}}

\newcommand{\pregzero}[1]{%
\begin{picture}(0.8,0.4)(-0.4,-0.4)
% label
\cell{0}{-0.4}{t}{#1}%
% dot
\cell{0}{0}{c}{\zmark}%
\end{picture}}

\mcm{\gzero}{1}{%
\ginitdims{0.8}{.6}%
\belowpic{#1}%
\sidespic{#1}%
\present{\pregzero{#1}}}

\newcommand{\pregone}[1]{%
\begin{picture}(5,0.4)(0,-0.2)%
% label
\cell{2.5}{0.2}{b}{#1}%
% arrow
\put(0,0){\vector(1,0){5}}%
\end{picture}}

\mcm{\gone}{1}{%
\ginitdims{5}{0.4}%
\abovepic{#1}%
\present{\pregone{#1}}}

\newcommand{\pregtwo}[3]{%
\begin{picture}(5,3.4)(0,-0.2)%
% labels
\cell{2.5}{3.2}{b}{#1}%
\cell{2.5}{-.2}{t}{#2}%
\cell{2.7}{1.5}{l}{#3}%
% arrows
\qbezier(0,1.5)(2.5,4.5)(5,1.5)%
\qbezier(0,1.5)(2.5,-1.5)(5,1.5)%
\put(5,1.5){\vector(1,-1){0}}%
\put(5,1.5){\vector(1,1){0}}%
\put(2.5,2.5){\vector(0,-1){2}}%
\end{picture}}

\mcm{\gtwo}{3}{%
\ginitdims{5}{3.4}%
\abovepic{#1}%
\belowpic{#2}%
\present{\pregtwo{#1}{#2}{#3}}}

\newcommand{\pregthree}[5]{%
\begin{picture}(5,5.4)(0,-1.2)%
% labels%
\cell{2.5}{4.2}{b}{#1}%
\cell{1.5}{1.7}{b}{#2}%
\cell{2.5}{-1.2}{t}{#3}%
\cell{2.7}{2.75}{l}{#4}%
\cell{2.7}{0.25}{l}{#5}%
% arrows%
\qbezier(0,1.5)(2.5,6.5)(5,1.5)%
\qbezier(0,1.5)(2.5,-3.5)(5,1.5)%
\put(0,1.5){\vector(1,0){5}}%
\put(2.5,3.5){\vector(0,-1){1.5}}%
\put(2.5,1){\vector(0,-1){1.5}}%
% arrowheads%
\put(5,1.5){\vector(1,-3){0}}%
\put(5,1.5){\vector(1,3){0}}%
\end{picture}}

\mcm{\gthree}{5}{%
\ginitdims{5}{5.4}%
\abovepic{#1}%
\belowpic{#3}%
\present{\pregthree{#1}{#2}{#3}{#4}{#5}}}

\newcommand{\pregfour}[7]{%
\begin{picture}(5,8.4)(0,-2.7)%
% labels%
\cell{2.5}{5.7}{b}{#1}%
\cell{1.5}{2.8}{b}{#2}%
\cell{1.5}{0.2}{t}{#3}%
\cell{2.5}{-2.7}{t}{#4}%
\cell{2.7}{4.25}{l}{#5}%
\cell{2.7}{1.5}{l}{#6}%
\cell{2.7}{-1.25}{l}{#7}%
% arrows%
\qbezier(0,1.5)(2.5,9.5)(5,1.5)%
\qbezier(0,1.5)(2.5,4)(5,1.5)%
\qbezier(0,1.5)(2.5,-1)(5,1.5)%
\qbezier(0,1.5)(2.5,-6.5)(5,1.5)%
\put(2.5,5.25){\vector(0,-1){2}}%
\put(2.5,2.5){\vector(0,-1){2}}%
\put(2.5,-0.25){\vector(0,-1){2}}%
% arrowheads%
\put(5,1.5){\vector(1,-4){0}}%
\put(5,1.5){\vector(4,-3){0}}%
\put(5,1.5){\vector(4,3){0}}%
\put(5,1.5){\vector(1,4){0}}%
\end{picture}}

\mcm{\gfour}{7}{%
\ginitdims{5}{8.4}%
\abovepic{#1}%
\belowpic{#4}%
\present{\pregfour{#1}{#2}{#3}{#4}{#5}{#6}{#7}}}

\newcommand{\pregthreecell}[5]{%
\begin{picture}(8,5)(-4,-2.5)%
% labels%
\cell{0}{2.5}{b}{#1}%
\cell{0}{-2.5}{t}{#2}%
\cell{-1.7}{0}{r}{#3}%
\cell{1.7}{0}{l}{#4}%
\cell{0}{0.2}{b}{#5}%
% arrows%
\qbezier(-4,0)(0,4.2)(4,0)%
\qbezier(-4,0)(0,-4.2)(4,0)%
\qbezier(-0.5,1.8)(-2.5,0)(-0.5,-1.8)%
\qbezier(0.5,1.8)(2.5,0)(0.5,-1.8)%
\put(-1,0){\vector(1,0){2}}%
% arrowheads%
\put(4,0){\vector(1,-1){0}}%
\put(4,0){\vector(1,1){0}}%
\put(-0.5,-1.8){\vector(1,-1){0}}%
\put(0.5,-1.8){\vector(-1,-1){0}}%
\end{picture}}

\mcm{\gthreecell}{5}{%
\ginitdims{8}{5}%
\abovepic{#1}%
\belowpic{#2}%
\present{\pregthreecell{#1}{#2}{#3}{#4}{#5}}}

%
% Special globular pictures
%

\newcommand{\pregthreecellu}{%
\begin{picture}(5,3.4)(-0.5,-0.2)%
% arrows%
\qbezier(-.5,1.5)(2,4.5)(4.5,1.5)%
\qbezier(-.5,1.5)(2,-1.5)(4.5,1.5)%
\qbezier(1.5,2.7)(0.5,1.5)(1.5,0.3)%
\qbezier(2.5,2.7)(3.5,1.5)(2.5,0.3)%
\put(1.3,1.5){\vector(1,0){1.4}}%
% arrowheads%
\put(4.5,1.5){\vector(1,-1){0}}%
\put(4.5,1.5){\vector(1,1){0}}%
\put(1.5,0.3){\vector(2,-3){0}}%
\put(2.5,0.3){\vector(-2,-3){0}}%
\end{picture}}

\mcm{\gthreecellu}{0}{%
\ginitdims{5}{3.4}%
\present{\pregthreecellu}}

\newcommand{\pregtwocentre}[3]{%
\begin{picture}(5,3.4)(0,-0.2)%
% labels
\cell{2.5}{3.2}{b}{#1}%
\cell{2.5}{-.2}{t}{#2}%
\cell{2.5}{1.5}{c}{#3}%
% arrows
\qbezier(0,1.5)(2.5,4.5)(5,1.5)%
\qbezier(0,1.5)(2.5,-1.5)(5,1.5)%
\put(5,1.5){\vector(1,-1){0}}%
\put(5,1.5){\vector(1,1){0}}%
\put(2.5,2.5){\vector(0,-1){2}}%
\end{picture}}

\mcm{\gtwocentre}{3}{%
\ginitdims{5}{3.4}%
\abovepic{#1}%
\belowpic{#2}%
\present{\pregtwocentre{#1}{#2}{#3}}}

\newcommand{\pregspecialone}[9]{%
\begin{picture}(8,8)(-4,-4)%
% labels%
\cell{0}{3.9}{b}{#1}%
\cell{-2}{-0.2}{t}{#2}%
\cell{0}{-3.9}{t}{#3}%
\cell{-1.5}{1.1}{r}{#4}%
\cell{0.2}{1.5}{l}{#5}%
\cell{1.5}{1.1}{l}{#6}%
\cell{0.2}{-2}{l}{#7}%
\cell{-0.9}{2.3}{b}{#8}%
\cell{0.9}{2.3}{b}{#9}%
% arrows%
\qbezier(-4,0)(0,8)(4,0)%
\qbezier(-4,0)(0,-8)(4,0)%
\qbezier(-0.5,3.4)(-3.5,2)(-0.5,0.6)%
\qbezier(0.5,3.4)(3.5,2)(0.5,0.6)%
\put(-4,0){\vector(1,0){8}}%
\put(0,3.4){\vector(0,-1){2.8}}%
\put(0,-0.8){\vector(0,-1){2.4}}%
\put(-1.5,2.2){\vector(1,0){1.2}}%
\put(0.3,2.2){\vector(1,0){1.2}}%
% arrowheads%
\put(4,0){\vector(1,-2){0}}%
\put(4,0){\vector(1,2){0}}%
\put(-0.5,0.6){\vector(2,-1){0}}%
\put(0.5,0.6){\vector(-2,-1){0}}%
\end{picture}}

\mcm{\gspecialone}{9}{%
\ginitdims{8}{8}%
\abovepic{#1}%
\belowpic{#3}%
\present{\pregspecialone{#1}{#2}{#3}{#4}{#5}{#6}{#7}{#8}{#9}}}

\newcommand{\pregspecialtwo}{%
\begin{picture}(5,3.4)(0,-0.2)%
% arrows
\qbezier(0,1.5)(2.5,4.5)(5,1.5)%
\qbezier(0,1.5)(2.5,-1.5)(5,1.5)%
\qbezier(1.7,2.5)(0,1.5)(1.7,0.5)%
\qbezier(3.3,2.5)(5,1.5)(3.3,0.5)%
% arrowheads
\put(5,1.5){\vector(1,-1){0}}%
\put(5,1.5){\vector(1,1){0}}%
\put(1.7,0.5){\vector(3,-2){0}}%
\put(3.3,0.5){\vector(-3,-2){0}}%
\put(2.5,2.5){\vector(0,-1){2}}%
\put(1.2,1.5){\vector(1,0){1}}%
\put(2.8,1.5){\vector(1,0){1}}%
\end{picture}}

\mcm{\gspecialtwo}{0}{%
\ginitdims{5}{3.4}%
\present{\pregspecialtwo}}

\newcommand{\pregspecialthree}{%
\begin{picture}(5,5.4)(0,-1.2)%
% arrows%
\qbezier(0,1.5)(2.5,6.5)(5,1.5)%
\qbezier(0,1.5)(2.5,-3.5)(5,1.5)%
\qbezier(2,3.5)(1,2.75)(2,2)%
\qbezier(3,3.5)(4,2.75)(3,2)%
\qbezier(2,1)(1,0.25)(2,-0.5)%
\qbezier(3,1)(4,0.25)(3,-0.5)%
\put(0,1.5){\vector(1,0){5}}%
\put(1.5,2.75){\vector(1,0){2}}%
\put(1.5,0.25){\vector(1,0){2}}%
% arrowheads%
\put(5,1.5){\vector(1,-3){0}}%
\put(5,1.5){\vector(1,3){0}}%
\put(2,2){\vector(1,-1){0}}%
\put(3,2){\vector(-1,-1){0}}%
\put(2,-0.5){\vector(1,-1){0}}%
\put(3,-0.5){\vector(-1,-1){0}}%
\end{picture}}

\mcm{\gspecialthree}{0}{%
\ginitdims{5}{5.4}%
\present{\pregspecialthree}}

\newcommand{\pregonew}[1]{%
\begin{picture}(8,0.4)(0,-0.2)%
% label
\cell{4}{0.2}{b}{#1}%
% arrow
\put(0,0){\vector(1,0){8}}%
\end{picture}}

\mcm{\gonew}{1}{%
\ginitdims{8}{0.4}%
\abovepic{#1}%
\present{\pregonew{#1}}}

\mcm{\gzersu}{0}{%
\gsinitdims{0}{.6}%
\present{\pregblw{}}}

\mcm{\gonesu}{0}{%
\gsinitdims{5}{0.4}%
\present{\pregone{}}}

\mcm{\gtwosu}{0}{%
\gsinitdims{5}{3.4}%
\present{\pregtwo{}{}{}}}

\mcm{\gthreesu}{0}{%
\gsinitdims{5}{5.4}%
\present{\pregthree{}{}{}{}{}}}

\mcm{\gfoursu}{0}{%
\gsinitdims{5}{8.4}%
\present{\pregfour{}{}{}{}{}{}{}}}

\newcommand{\precone}[1]{%
\begin{picture}(4.2,0.4)(-0.3,-0.2)%
% label
\cell{1.8}{0.2}{b}{#1}%
% arrow
\put(0,0){\vector(1,0){3.6}}%
\end{picture}}

\mcm{\cone}{1}{%
\cinitdims{4.2}{0.4}%
\abovepic{#1}%
\present{\precone{#1}}}

\mcm{\gfstsu}{0}{%
\gsinitdims{0.5}{0.4}%
\presentr{\pregfst{}}}

\mcm{\glstsu}{0}{%
\gsinitdims{0.5}{0.4}%
\presentl{\preglst{}}}

% Cellular cells with double and triple arrows (drawn full-length, not with
% LaTeX characters)

\newcommand{\prectwodbl}[3]%
{\begin{picture}(4.2,3.4)(-0.1,-0.2)%
% labels%
\cell{2}{3.2}{b}{#1}%
\cell{2}{-0.2}{t}{#2}%
\cell{2.3}{1.5}{l}{#3}%
% arrows%
\qbezier(0,2)(2,4)(4,2)%
\qbezier(0,1)(2,-1)(4,1)%
\put(4,2){\vector(1,-1){0}}%
\put(4,1){\vector(1,1){0}}%
\put(1.9,2.5){\line(0,-1){1.8}}%
\put(2.1,2.5){\line(0,-1){1.8}}%
\cell{2.01}{0.4}{b}{\vee}%
\end{picture}}

\mcm{\ctwodbl}{3}{%
\cinitdims{4.2}{3.4}%
\abovepic{#1}%
\belowpic{#2}%
\present{\prectwodbl{#1}{#2}{#3}}}

\newcommand{\precthreedbl}[5]{%
\begin{picture}(4.2,5.4)(-0.1,-0.2)%
% labels%
\cell{2}{5.2}{b}{#1}%
\cell{1}{2.7}{b}{#2}%
\cell{2}{-.2}{t}{#3}%
\cell{2.3}{3.75}{l}{#4}%
\cell{2.3}{1.25}{l}{#5}%
% arrows%
\qbezier(0,3)(2,7)(4,3)%
\qbezier(0,2)(2,-2)(4,2)%
\put(0,2.5){\vector(1,0){4}}%
\put(1.9,4.5){\line(0,-1){1.3}}%
\put(2.1,4.5){\line(0,-1){1.3}}%
\cell{2.01}{2.9}{b}{\vee}%
\put(1.9,2){\line(0,-1){1.3}}%
\put(2.1,2){\line(0,-1){1.3}}%
\cell{2.01}{0.4}{b}{\vee}%
% arrowheads%
\put(4,3){\vector(1,-3){0}}%
\put(4,2){\vector(1,3){0}}%
\end{picture}}

\mcm{\cthreedbl}{5}{%
\cinitdims{4.2}{5.4}%
\abovepic{#1}%
\belowpic{#3}%
\present{\precthreedbl{#1}{#2}{#3}{#4}{#5}}}

\newcommand{\precthreecelltrp}[5]{%
\begin{picture}(8.2,5)(-4.1,-2.5)%
% labels%
\cell{0}{2.5}{b}{#1}%
\cell{0}{-2.5}{t}{#2}%
\cell{-1.8}{0}{r}{#3}%
\cell{1.8}{0}{l}{#4}%
\cell{0}{0.3}{b}{#5}%
% arrows%
\qbezier(-4,0.5)(0,4)(4,0.5)%
\qbezier(-4,-0.5)(0,-4)(4,-0.5)%
\qbezier(-0.6,2)(-2.6,0)(-0.6,-2)%
\qbezier(-0.4,2)(-2.4,0)(-0.5,-1.9)%
\cell{-0.6}{-2}{b}{\lrcorner}%
\qbezier(0.4,2)(2.4,0)(0.5,-1.9)%
\qbezier(0.6,2)(2.6,0)(0.6,-2)%
\cell{0.65}{-2}{b}{\llcorner}%
\put(-1,0.15){\line(1,0){1.7}}%
\put(-1,0){\line(1,0){2}}%
\put(-1,-0.15){\line(1,0){1.7}}%
\cell{1.15}{0}{r}{>}%
% arrowheads%
\put(4,0.5){\vector(1,-1){0}}%
\put(4,-0.5){\vector(1,1){0}}%
\end{picture}}

\mcm{\cthreecelltrp}{5}{%
\cinitdims{8.2}{5}%
\abovepic{#1}%
\belowpic{#2}%
\present{\precthreecelltrp{#1}{#2}{#3}{#4}{#5}}}

%
% CELLULAR PICTURES
%

\newcommand{\prectwo}[3]%
{\begin{picture}(4.2,3.4)(-0.1,-0.2)%
% labels%
\cell{2}{3.2}{b}{#1}%
\cell{2}{-0.2}{t}{#2}%
\cell{2.2}{1.5}{l}{#3}%
% arrows%
\qbezier(0,2)(2,4)(4,2)%
\qbezier(0,1)(2,-1)(4,1)%
\put(4,2){\vector(1,-1){0}}%
\put(4,1){\vector(1,1){0}}%
\put(2,2.5){\vector(0,-1){2}}%
\end{picture}}

\mcm{\ctwo}{3}{%
\cinitdims{4.2}{3.4}%
\abovepic{#1}%
\belowpic{#2}%
\present{\prectwo{#1}{#2}{#3}}}

\newcommand{\precthree}[5]{%
\begin{picture}(4.2,5.4)(-0.1,-0.2)%
% labels%
\cell{2}{5.2}{b}{#1}%
\cell{1}{2.7}{b}{#2}%
\cell{2}{-.2}{t}{#3}%
\cell{2.2}{3.75}{l}{#4}%
\cell{2.2}{1.25}{l}{#5}%
% arrows%
\qbezier(0,3)(2,7)(4,3)%
\qbezier(0,2)(2,-2)(4,2)%
\put(0,2.5){\vector(1,0){4}}%
\put(2,4.5){\vector(0,-1){1.5}}%
\put(2,2){\vector(0,-1){1.5}}%
% arrowheads%
\put(4,3){\vector(1,-3){0}}%
\put(4,2){\vector(1,3){0}}%
\end{picture}}

\mcm{\cthree}{5}{%
\cinitdims{4.2}{5.4}%
\abovepic{#1}%
\belowpic{#3}%
\present{\precthree{#1}{#2}{#3}{#4}{#5}}}

%
% Special cellular pictures (horizontal)
%

\newcommand{\prectwoop}[3]%
{\begin{picture}(4.2,3.4)(-0.1,-0.2)%
% labels%
\cell{2}{3.2}{b}{#1}%
\cell{2}{-0.2}{t}{#2}%
\cell{2.2}{1.5}{l}{#3}%
% arrows%
\qbezier(0,2)(2,4)(4,2)%
\qbezier(0,1)(2,-1)(4,1)%
\put(0,2){\vector(-1,-1){0}}%
\put(0,1){\vector(-1,1){0}}%
\put(2,2.5){\vector(0,-1){2}}%
\end{picture}}

\mcm{\ctwoop}{3}{%
\cinitdims{4.2}{3.4}%
\abovepic{#1}%
\belowpic{#2}%
\present{\prectwoop{#1}{#2}{#3}}}

\newcommand{\prectwopar}[4]{%
\begin{picture}(4.2,3.4)(-0.1,-0.2)%
% labels%
\cell{2}{3.2}{b}{#1}%
\cell{2}{-0.2}{t}{#2}%
\cell{1.6}{1.5}{r}{#3}%
\cell{2.4}{1.5}{l}{#4}%
% arrows%
\qbezier(0,2)(2,4)(4,2)%
\qbezier(0,1)(2,-1)(4,1)%
\put(4,2){\vector(1,-1){0}}%
\put(4,1){\vector(1,1){0}}%
\put(1.8,2.5){\vector(0,-1){2}}%
\put(2.2,2.5){\vector(0,-1){2}}%
\end{picture}}

\mcm{\ctwopar}{4}{%
\cinitdims{4.2}{3.4}%
\abovepic{#1}%
\belowpic{#2}%
\present{\prectwopar{#1}{#2}{#3}{#4}}}

\newcommand{\precthreein}[5]{%
\begin{picture}(4.2,5.4)(-0.1,-0.2)%
% labels%
\cell{2}{5.2}{b}{#1}%
\cell{1}{2.7}{b}{#2}%
\cell{2}{-.2}{t}{#3}%
\cell{2.2}{3.75}{l}{#4}%
\cell{2.2}{1.25}{l}{#5}%
% arrows%
\qbezier(0,3)(2,7)(4,3)%
\qbezier(0,2)(2,-2)(4,2)%
\put(0,2.5){\vector(1,0){4}}%
\put(2,4.5){\vector(0,-1){1.5}}%
\put(2,0.5){\vector(0,1){1.5}}%
% arrowheads%
\put(4,3){\vector(1,-3){0}}%
\put(4,2){\vector(1,3){0}}%
\end{picture}}

\mcm{\cthreein}{5}{%
\cinitdims{4.2}{5.4}%
\abovepic{#1}%
\belowpic{#3}%
\present{\precthreein{#1}{#2}{#3}{#4}{#5}}}

\newcommand{\precthreecell}[5]{%
\begin{picture}(8.2,5)(-4.1,-2.5)%
% labels%
\cell{0}{2.5}{b}{#1}%
\cell{0}{-2.5}{t}{#2}%
\cell{-1.7}{0}{r}{#3}%
\cell{1.7}{0}{l}{#4}%
\cell{0}{0.2}{b}{#5}%
% arrows%
\qbezier(-4,0.5)(0,4)(4,0.5)%
\qbezier(-4,-0.5)(0,-4)(4,-0.5)%
\qbezier(-0.5,2)(-2.5,0)(-0.5,-2)%
\qbezier(0.5,2)(2.5,0)(0.5,-2)%
\put(-1,0){\vector(1,0){2}}%
% arrowheads%
\put(4,0.5){\vector(1,-1){0}}%
\put(4,-0.5){\vector(1,1){0}}%
\put(-0.5,-2){\vector(1,-1){0}}%
\put(0.5,-2){\vector(-1,-1){0}}%
\end{picture}}

\mcm{\cthreecell}{5}{%
\cinitdims{8.2}{5}%
\abovepic{#1}%
\belowpic{#2}%
\present{\precthreecell{#1}{#2}{#3}{#4}{#5}}}

\newcommand{\precthreecellpar}[6]{%
\begin{picture}(8.2,5)(-4.1,-2.5)%
% labels%
\cell{0}{2.5}{b}{#1}%
\cell{0}{-2.5}{t}{#2}%
\cell{-1.7}{0}{r}{#3}%
\cell{1.7}{0}{l}{#4}%
\cell{0}{0.4}{b}{#5}%
\cell{0}{-0.4}{t}{#6}%
% arrows%
\qbezier(-4,0.5)(0,4)(4,0.5)%
\qbezier(-4,-0.5)(0,-4)(4,-0.5)%
\qbezier(-0.5,2)(-2.5,0)(-0.5,-2)%
\qbezier(0.5,2)(2.5,0)(0.5,-2)%
\put(-1,0.2){\vector(1,0){2}}%
\put(-1,-0.2){\vector(1,0){2}}%
% arrowheads%
\put(4,0.5){\vector(1,-1){0}}%
\put(4,-0.5){\vector(1,1){0}}%
\put(-0.5,-2){\vector(1,-1){0}}%
\put(0.5,-2){\vector(-1,-1){0}}%
\end{picture}}

\mcm{\cthreecellpar}{6}{%
\cinitdims{8.2}{5}%
\abovepic{#1}%
\belowpic{#2}%
\present{\precthreecellpar{#1}{#2}{#3}{#4}{#5}{#6}}}

%
% Special cellular pictures (vertical)
%

\newcommand{\prectwov}[5]{%
\begin{picture}(3.4,4.2)(0.8,0.9)%
% labels%
\cell{2.5}{5.1}{b}{#1}%
\cell{2.5}{0.9}{t}{#2}%
\cell{0.8}{3}{r}{#3}%
\cell{4.2}{3}{l}{#4}%
\cell{2.5}{3.2}{b}{#5}%
% arrows%
\qbezier(2,5)(0,3)(2,1)%
\qbezier(3,5)(5,3)(3,1)%
\put(2,1){\vector(1,-1){0}}%
\put(3,1){\vector(-1,-1){0}}%
\put(1.5,3){\vector(1,0){2}}%
\end{picture}}

\mcm{\ctwov}{5}{%
\cinitdims{3.4}{4.2}%
\abovepic{#1}%
\belowpic{#2}%
\sidespic{#3}%
\sidespic{#4}%
\present{\prectwov{#1}{#2}{#3}{#4}{#5}}}

\newcommand{\precthreecellv}[7]{%
\begin{picture}(5,8.2)(0.5,-1.6)%
% labels%
\cell{3}{6.6}{b}{#1}%
\cell{3}{-1.6}{t}{#2}%
\cell{0.5}{2.5}{r}{#3}%
\cell{5.5}{2.5}{l}{#4}%
\cell{3}{4.2}{b}{#5}%
\cell{3}{0.8}{t}{#6}%
\cell{3.2}{2.5}{l}{#7}%
% arrows%
\qbezier(3.5,6.5)(7,2.5)(3.5,-1.5)%
\qbezier(2.5,6.5)(-1,2.5)(2.5,-1.5)%
\put(2.5,-1.5){\vector(1,-1){0}}%
\put(3.5,-1.5){\vector(-1,-1){0}}%
\qbezier(1,3)(3,5)(5,3)%
\qbezier(1,2)(3,0)(5,2)%
\put(5,3){\vector(1,-1){0}}%
\put(5,2){\vector(1,1){0}}%
\put(3,3.5){\vector(0,-1){2}}%
\end{picture}}

\mcm{\cthreecellv}{7}{%
\cinitdims{5}{8.2}%
\abovepic{#1}%
\belowpic{#2}%
\sidespic{#3}%
\sidespic{#4}%
\present{\precthreecellv{#1}{#2}{#3}{#4}{#5}{#6}{#7}}}

%%%%%%%%%%%%%%%%%%%%%%%%%%%%%%%%%%%%%%%%%%%%%%%%%%%%%%%%%%%%%%%%%%%%%%%%%%%%
%
% OPETOPES
%

\newcommand{\pretopez}[2]{%
\begin{picture}(2.6,2.3)(-1.3,-2.2)%
% labels
\cell{0}{-2.2}{t}{#1}%
\cell{0}{-1.2}{c}{#2}%
% dot
% \cell{0}{0}{c}{\zmark}%
% edge
\qbezier(0,0)(-2,-2)(0,-2)%
\qbezier(0,0)(2,-2)(0,-2)%
% arrowhead
\put(0,0){\vector(-1,1){0}}%
\end{picture}}

\mcm{\topez}{2}{%
\ginitdims{2.6}{2.3}%
\belowpic{#1}%
\present{\pretopez{#1}{#2}}}

\newcommand{\pretopea}[3]{%
\begin{picture}(4,1.9)(-2,-0,2)%
% labels
\cell{0}{1.7}{b}{#1}%
\cell{0}{-0.2}{t}{#2}%
\cell{0}{0.7}{c}{#3}%
% edges
\qbezier(-2,0)(0,3)(2,0)%
\put(-2,0){\vector(1,0){4}}%
% arrowhead
\put(2,0){\vector(2,-3){0}}%
\end{picture}}

\mcm{\topea}{3}{%
\ginitdims{4}{1.9}%
\abovepic{#1}%
\belowpic{#2}%
\present{\pretopea{#1}{#2}{#3}}}

\newcommand{\pretopeb}[4]{%
\begin{picture}(4,2.2)(-2,-0.2)%
% labels
\cell{-1.1}{1}{br}{#1}%
\cell{1.1}{1}{bl}{#2}%
\cell{0}{-0.2}{t}{#3}%
\cell{0}{0.8}{c}{#4}%
% edges
\put(-2,0){\vector(1,1){2}}%
\put(0,2){\vector(1,-1){2}}%
\put(-2,0){\vector(1,0){4}}%
\end{picture}}

\mcm{\topeb}{4}{%
\ginitdims{4}{2.2}%
\belowpic{#3}%
\present{\pretopeb{#1}{#2}{#3}{#4}}}

\newcommand{\pretopec}[5]{%
\begin{picture}(4,2.2)(-2,-0.2)%
% labels
\cell{-1.8}{1}{br}{#1}%
\cell{0}{2.2}{b}{#2}%
\cell{1.8}{1}{bl}{#3}%
\cell{0}{-0.2}{t}{#4}%
\cell{0}{0.8}{c}{#5}%
% edges
\put(-2,0){\vector(1,2){1}}%
\put(-1,2){\vector(1,0){2}}%
\put(1,2){\vector(1,-2){1}}%
\put(-2,0){\vector(1,0){4}}%
\end{picture}}

\mcm{\topec}{5}{%
\ginitdims{4}{2.2}%
\sidespic{#1}%
\abovepic{#2}%
\sidespic{#3}%
\belowpic{#4}%
\present{\pretopec{#1}{#2}{#3}{#4}{#5}}}

\newcommand{\pretoped}[6]{%
\begin{picture}(4,2.5)(-2,-0.2)%
% labels
\cell{-2}{0.6}{br}{#1}%
\cell{-0.7}{2.2}{br}{#2}%
\cell{0.7}{2.2}{bl}{#3}%
\cell{2}{0.6}{bl}{#4}%
\cell{0}{-0.2}{t}{#5}%
\cell{0}{0.8}{c}{#6}%
% edges
\put(-2,0){\vector(1,3){0.5}}%
\put(-1.5,1.5){\vector(3,2){1.5}}%
\put(0,2.5){\vector(3,-2){1.5}}%
\put(1.5,1.5){\vector(1,-3){0.5}}%
\put(-2,0){\vector(1,0){4}}%
\end{picture}}

\mcm{\toped}{6}{%
\ginitdims{4}{2.5}%
\sidespic{#1}%
\abovepic{#2}%
\abovepic{#3}%
\sidespic{#4}%
\belowpic{#5}%
\present{\pretoped{#1}{#2}{#3}{#4}{#5}{#6}}}

\newcommand{\pretopeq}[5]{%
\begin{picture}(4,2.5)(-2,-0.2)%
% labels
\cell{-2}{0.6}{br}{#1}%
\cell{-1}{2.2}{br}{#2}%
\cell{2}{0.6}{bl}{#3}%
\cell{0}{-0.2}{t}{#4}%
\cell{0}{0.8}{c}{#5}%
% edges
\put(-2,0){\vector(1,3){0.5}}%
\put(-1.5,1.5){\vector(1,1){1}}%
\cell{0.9}{2.3}{c}{\ddots}
\put(1.5,1.5){\vector(1,-3){0.5}}%
\put(-2,0){\vector(1,0){4}}%
\end{picture}}

\mcm{\topeq}{5}{%
\ginitdims{4}{2.5}%
\sidespic{#1}%
\abovepic{#2}%
\sidespic{#3}%
\belowpic{#4}%
\present{\pretopeq{#1}{#2}{#3}{#4}{#5}}}

\newcommand{\pretopebase}[1]{%
\begin{picture}(4,0.4)(0,-0.2)%
% label
\cell{2}{0.2}{b}{#1}%
% arrow
\put(0,0){\vector(1,0){4}}%
\end{picture}}

\mcm{\topebase}{1}{%
\ginitdims{4}{0.4}%
\abovepic{#1}%
\present{\pretopebase{#1}}}

%%%%%%%%%%%%%%%%%%%%%%%%%%%%%%%%%%%%%%%%%%%%%%%%%%%%%%%%%%%%%%%%%%%%%%%%%%%%
%
% OPETOPES STRIPPED OF ARROWHEADS
%

\newcommand{\pretopezs}[2]{%
\begin{picture}(2.6,2.3)(-1.3,-2.2)%
% labels
\cell{0}{-2.2}{t}{#1}%
\cell{0}{-1.2}{c}{#2}%
% dot
% \cell{0}{0}{c}{\zmark}%
% edge
\qbezier(0,0)(-2,-2)(0,-2)%
\qbezier(0,0)(2,-2)(0,-2)%
\end{picture}}

\mcm{\topezs}{2}{%
\ginitdims{2.6}{2.3}%
\belowpic{#1}%
\present{\pretopezs{#1}{#2}}}

\newcommand{\pretopeas}[3]{%
\begin{picture}(4,1.9)(-2,-0,2)%
% labels
\cell{0}{1.7}{b}{#1}%
\cell{0}{-0.2}{t}{#2}%
\cell{0}{0.7}{c}{#3}%
% edges
\qbezier(-2,0)(0,3)(2,0)%
\put(-2,0){\line(1,0){4}}%
\end{picture}}

\mcm{\topeas}{3}{%
\ginitdims{4}{1.9}%
\abovepic{#1}%
\belowpic{#2}%
\present{\pretopeas{#1}{#2}{#3}}}

\newcommand{\pretopebs}[4]{%
\begin{picture}(4,2.2)(-2,-0.2)%
% labels
\cell{-1.1}{1}{br}{#1}%
\cell{1.1}{1}{bl}{#2}%
\cell{0}{-0.2}{t}{#3}%
\cell{0}{0.8}{c}{#4}%
% edges
\put(-2,0){\line(1,1){2}}%
\put(0,2){\line(1,-1){2}}%
\put(-2,0){\line(1,0){4}}%
\end{picture}}

\mcm{\topebs}{4}{%
\ginitdims{4}{2.2}%
\belowpic{#3}%
\present{\pretopebs{#1}{#2}{#3}{#4}}}

\newcommand{\pretopecs}[5]{%
\begin{picture}(4,2.2)(-2,-0.2)%
% labels
\cell{-1.8}{1}{br}{#1}%
\cell{0}{2.2}{b}{#2}%
\cell{1.8}{1}{bl}{#3}%
\cell{0}{-0.2}{t}{#4}%
\cell{0}{0.8}{c}{#5}%
% edges
\put(-2,0){\line(1,2){1}}%
\put(-1,2){\line(1,0){2}}%
\put(1,2){\line(1,-2){1}}%
\put(-2,0){\line(1,0){4}}%
\end{picture}}

\mcm{\topecs}{5}{%
\ginitdims{4}{2.2}%
\sidespic{#1}%
\abovepic{#2}%
\sidespic{#3}%
\belowpic{#4}%
\present{\pretopecs{#1}{#2}{#3}{#4}{#5}}}

\newcommand{\pretopeds}[6]{%
\begin{picture}(4,2.5)(-2,-0.2)%
% labels
\cell{-2}{0.6}{br}{#1}%
\cell{-0.7}{2.2}{br}{#2}%
\cell{0.7}{2.2}{bl}{#3}%
\cell{2}{0.6}{bl}{#4}%
\cell{0}{-0.2}{t}{#5}%
\cell{0}{0.8}{c}{#6}%
% edges
\put(-2,0){\line(1,3){0.5}}%
\put(-1.5,1.5){\line(3,2){1.5}}%
\put(0,2.5){\line(3,-2){1.5}}%
\put(1.5,1.5){\line(1,-3){0.5}}%
\put(-2,0){\line(1,0){4}}%
\end{picture}}

\mcm{\topeds}{6}{%
\ginitdims{4}{2.5}%
\sidespic{#1}%
\abovepic{#2}%
\abovepic{#3}%
\sidespic{#4}%
\belowpic{#5}%
\present{\pretopeds{#1}{#2}{#3}{#4}{#5}{#6}}}

\newcommand{\pretopeqs}[5]{%
\begin{picture}(4,2.5)(-2,-0.2)%
% labels
\cell{-2}{0.6}{br}{#1}%
\cell{-1}{2.2}{br}{#2}%
\cell{2}{0.6}{bl}{#3}%
\cell{0}{-0.2}{t}{#4}%
\cell{0}{0.8}{c}{#5}%
% edges
\put(-2,0){\line(1,3){0.5}}%
\put(-1.5,1.5){\line(1,1){1}}%
\cell{0.9}{2.3}{c}{\ddots}
\put(1.5,1.5){\line(1,-3){0.5}}%
\put(-2,0){\line(1,0){4}}%
\end{picture}}

\mcm{\topeqs}{5}{%
\ginitdims{4}{2.5}%
\sidespic{#1}%
\abovepic{#2}%
\sidespic{#3}%
\belowpic{#4}%
\present{\pretopeqs{#1}{#2}{#3}{#4}{#5}}}

\newcommand{\pretopebases}[1]{%
\begin{picture}(4,0.4)(0,-0.2)%
% label
\cell{2}{0.2}{b}{#1}%
% arrow
\put(0,0){\line(1,0){4}}%
\end{picture}}

\mcm{\topebases}{1}{%
\ginitdims{4}{0.4}%
\abovepic{#1}%
\present{\pretopebases{#1}}}

% Notes and changes
%
% ctwocellv has been changed to ctwov

% This is a new one, adapted from gfour and used in defn.tex (thesis)

\newcommand{\pregdots}[6]{%
\begin{picture}(5,8.4)(0,-2.7)%
% labels%
\cell{2.5}{5.7}{b}{#1}%
\cell{1.5}{2.8}{b}{#2}%
\cell{1.5}{0.2}{t}{#3}%
\cell{2.5}{-2.7}{t}{#4}%
\cell{2.7}{4.25}{l}{#5}%
\cell{2.7}{-1.25}{l}{#6}%
% arrows%
\qbezier(0,1.5)(2.5,9.5)(5,1.5)%
\qbezier(0,1.5)(2.5,4)(5,1.5)%
\qbezier(0,1.5)(2.5,-1)(5,1.5)%
\qbezier(0,1.5)(2.5,-6.5)(5,1.5)%
\put(2.5,5.25){\vector(0,-1){2}}%
\put(2.5,-0.25){\vector(0,-1){2}}%
% vertical ellipsis%
\cell{2.5}{1.7}{c}{\vdots}%
% arrowheads%
\put(5,1.5){\vector(1,-4){0}}%
\put(5,1.5){\vector(4,-3){0}}%
\put(5,1.5){\vector(4,3){0}}%
\put(5,1.5){\vector(1,4){0}}%
\end{picture}}

\mcm{\gdots}{6}{%
\ginitdims{5}{8.4}%
\abovepic{#1}%
\belowpic{#4}%
\present{\pregdots{#1}{#2}{#3}{#4}{#5}{#6}}}

%%%%%%%%%%%%%%%%%%%%%%%%%%%%%%%%%%%%%%%%%%%%%%%%%%%%%%%%%%%%%%%%%%%%%%%%%

% Tree shapes

% generic tree of height 1
%
\newcommand{\GenericOne}[3]
{\begin{tree}
\nl{#1} &   &\nl{#2}&   &   &\cdots & & &\nl{#3}    \\
    &\rt{4} &   &\rt{2} &   &   & &\lt{4} & \\
    &   &   &   &\node  &   & & &       \\
\end{tree}}

% empty-headed tree of height 1
%
\newcommand{\EmptyOne}
{\begin{tree}
\enode  \\
\dn \\
\node   \\
\end{tree}}

% 0,5,3,1 is 'formation' of this tree; there are 5 leaves
%
\newcommand{\Oak}[5]
{\begin{tree}
 & & & &\enode& & & & & & & & \\
 & & & &\dn& & & & & & & & \\
\nl{#1}& &\nl{#2}& &\node& & & &\nl{#4}& & & &\nl{#5}\\
 &\rt{2}&\dn&\lt{2}& & & & & &\rt{2}& &\lt{2}& \\
 & &\node& & & &\nl{#3}& & & &\node& & \\
 & & &\rt{4}& & &\dn& & &\lt{4}& & & \\
 & & & & & &\node& & & & & & \\
\end{tree}}

% 0,2,2,1; 2
%
\newcommand{\Pear}[3]
{\begin{tree}
    &   &   &   &\enode &&  &   &   \\
    &   &   &   &\dn    &&  &   &   \\
\nl{#1}&    &   &   &\node  &&  &   &   \\
    &\rt{2} &   &\lt{2} &   &&  &   &   \\
    &   &\node  &   &   &&  &   &\nl{#2}\\
    &   &   &\rt{4} &   &&  &\lt{2} &   \\
    &   &   &   &   &&\nll{#3}& &   \\
\end{tree}}

% 1,2,1; 2
%
\newcommand{\Orange}[3]
{\begin{tree}
    &   &   &   &\nl{#2}\\
    &   &   &   &\dn    \\
\nl{#1}&    &   &   &\node  \\
    &\rt{2} &   &\lt{2} &   \\
    &   &\nll{#3}&  &   \\
\end{tree}}

% 0,3,1,1; 2
%
\newcommand{\Apple}[3]
{\begin{tree}
    &   &\enode &   &   \\
    &   &\dn    &   &   \\
\nl{#1} &   &\node  &   &\nl{#2}\\
    &\rt{2} &\dn    &\lt{2} &   \\
    &   &\node  &   &   \\
    &   &\dn    &   &   \\
    &   &\nll{#3}&  &   \\
\end{tree}}

% This is orange and apple substituted into pear
% 1,5,3,2,1; 4
%
\newcommand{\MixedFruit}[5]
{\begin{tree}
 & & & &\nl{#2}& & & & & & &\enode& & \\
 & & & &\dn    & & & & & & &\dn   & & \\
\nl{#1}& & & &\node& & &\enode& &\nl{#3}& &\node& &\nl{#4}\\
 &\rt{2}& &\lt{2}& & & &\dn& & &\rt{2}&\dn&\lt{2}& \\
 & &\node& & & & &\node& & & &\node& & \\
 & & &\rt{3}& & &\lt{2}& & & & &\dn& & \\
 & & & & &\node& & & & & &\node& & \\
 & & & & & &\rt{3}& & & &\lt{3}& & & \\
 & & & & & & & &\nll{#5}& & & & & \\
\end{tree}}

%%%%%%%%%%%%%%%%%%%%%%%%%%%%%%%%%%%%%%%%%%%%%%%%%%%%%%%%%%%%%%%%%%%%%%%%%
%
% Unlabelled trees

\newcommand{\treedc}{
\begin{tree}
\node &\node            &\node &      &      &\node \\
      &\rt{1} \dn \lt{1}&      &      &      & \dn  \\
      &\node            &      &\node &      &\node \\
      &                 &\rt{2}&\dn   &\lt{2}&      \\
      &                 &      &\node &      &      \\
\end{tree}}

\newcommand{\treedcbang}{
\begin{tree}
\node &                 &\node &      &      &\node \\
      &\rt{1}     \lt{1}&      &      &      & \dn  \\
      &\node            &      &\node &      &\node \\
      &                 &\rt{2}&\dn   &\lt{2}&      \\
      &                 &      &\node &      &      \\
\end{tree}}

\newcommand{\treec}{
\begin{tree}
\node   &   &\node  &   &\node  \\
    &\rt{2} &\dn    &\lt{2} &   \\
    &   &\node  &   &   \\
\end{tree}}

\newcommand{\treecc}{
\begin{tree}
    &   &   &\node  &\node          &\node  \\
    &   &   &   &\rt{1} \dn \lt{1}  &   \\
\node   &   &\node  &   &\node          &   \\
    &\rt{2} &\dn    &\lt{2} &           &   \\
    &   &\node  &   &           &   \\
\end{tree}}

\newcommand{\treeccbang}{
\begin{tree}
    &   &   &\node  &           &\node  \\
    &   &   &   &\rt{1}  \lt{1}     &   \\
\node   &   &\node  &   &\node          &   \\
    &\rt{2} &\dn    &\lt{2} &           &   \\
    &   &\node  &   &           &   \\
\end{tree}}

\newcommand{\treegc}{
\begin{tree}
\node &\node    &\node  &   &\node  &\node  &\node      &\node  \\
&\rt{1} \dn \lt{1}& &   &   &\rt{1}\dn\lt{1}&\lt{2} &   \\
    &\node  &   &\node  &   &\node  &       &   \\
    &   &\rt{2} &\dn    &\lt{2} &   &       &   \\
    &   &   &\node  &   &   &       &   \\
\end{tree}}

\newcommand{\treegcbang}{
\begin{tree}
\node &     &\node  &   &\node  &\node  &\node      &   \\
&\rt{1}  \lt{1} &   &   &   &\rt{1}\dn\lt{1}&   &   \\
    &\node  &   &\node  &   &\node  &       &   \\
    &   &\rt{2} &\dn    &\lt{2} &   &       &   \\
    &   &   &\node  &   &   &       &   \\
\end{tree}}

\newcommand{\treebaa}{
\begin{tree}
\node   &       &\node  \\
    &\rt{1}\lt{1}   &   \\
    &\node      &   \\
    &\dn        &   \\
    &\node      &   \\
    &\dn        &   \\
    &\node      &   \\
\end{tree}}

\newcommand{\treeba}{
\begin{tree}
\node   &       &\node  \\
    &\rt{1}\lt{1}   &   \\
    &\node      &   \\
    &\dn        &   \\
    &\node      &   \\
\end{tree}}

\newcommand{\treebcb}{
\begin{tree}
\node   &   &\node  &   &   &   \\
    &\rt{1}\lt{1}&  &   &   &   \\
    &\node  &   &\node  &   &\node  \\
    &\dn    &   &   &\rt{1}\lt{1}&  \\
    &\node  &   &   &\node  &   \\
    &   &\rt{1} &\lt{2} &   &   \\
    &   &\node  &   &   &   \\
\end{tree}}

\newcommand{\treeb}{
\begin{tree}
\node   &       &\node  \\
    &\rt{1}\lt{1}   &   \\
    &\node      &   \\
\end{tree}}

\newcommand{\treedb}{
\begin{tree}
\node   &   &   &\node  &\node      &\node  \\
\dn &   &   &   &\rt{1}\dn\lt{1}&   \\
\node   &   &   &   &\node      &   \\
    &\rt{2} &   &\lt{2} &       &   \\
    &   &\node  &   &       &   \\
\end{tree}}

\newcommand{\treebb}{
\begin{tree}
    &   &   &\node  &       &\node  \\
    &   &   &   &\rt{1}\lt{1}   &   \\
\node   &   &   &   &\node      &   \\
    &\rt{2} &   &\lt{2} &       &   \\
    &   &\node  &   &       &   \\
\end{tree}}

\newcommand{\treefg}{
\begin{tree}
\node   &\node  &\node  &\node  &   &\node  &   &\node  &
    &   &   &   &   \\
\dn &   &\rt{1}\dn\lt{1}&&  &   &\rt{1}\lt{1}&  &
    &   &   &   &   \\
\node   &   &\node  &   &\node  &   &\node  &   &
\node   &   &\node  &   &\node  \\
    &\rt{6} &   &\rt{4} &   &\rt{2} &\dn    &\lt{2} &
    &\lt{4} &   &\lt{6} &   \\
    &   &   &   &   &   &\node  &   &
    &   &   &   &   \\
\end{tree}}

\newcommand{\treeec}{
\begin{tree}
    &   &\node  &\node      &\node  &\node  &   &\node  \\
    &   &   &\rt{1} \dn \lt{1}& &   &\rt{1}\lt{1}&  \\
\node   &   &   &\node      &   &   &\node  &   \\
    &\rt{3} &   &\dn        &   &\lt{3} &   &   \\
    &   &   &\node      &   &   &   &   \\
\end{tree}}

\newcommand{\treeaaq}{
\begin{tree}
\node   \\
\dn \\
\node   \\
\dn \\
\node   \\
\vdots  \\
\node   \\
\dn \\
\node   \\
\end{tree}}

\newcommand{\treeq}{
\begin{tree}
\node   &\node  &   &\ldots &   &\node  \\
    &\rt{2} &\rt{1} &   &\lt{3} &   \\
    &   &\node  &   &   &   \\
\end{tree}}

\newcommand{\treea}{
\begin{tree}
\node   \\
\dn \\
\node   \\
\end{tree}}

%%%%%%%%%%%%%%%%%%%%%%%%%%%%%%%%%%%%%%%%%%%%%%%%%%%%%%%%%%%%%%%%%%%%%%%%%
%
% Transistor section

\newlength{\volt}
\setlength{\volt}{1ex}

% A transistor; used like `transistor{a}{s1}{s2}{sn}{s}'
%
\newcommand{\transistor}[5]
{\setlength{\unitlength}{1\volt}
\begin{picture}(18,12)(-5,-6)
% main triangle
\put(0,6){\line(0,-1){12}} \put(0,-6){\line(3,2){9}}
\put(0,6){\line(3,-2){9}}
% legs
\put(9,0){\line(1,0){2}} \put(-2,4){\line(1,0){2}}
\put(-2,2){\line(1,0){2}} \put(-2,-4){\line(1,0){2}}
% leg-labels
\put(12,-0.5){\ensuremath{#5}} \put(-5,3.5){\ensuremath{#2}}
\put(-5,1.5){\ensuremath{#3}} \put(-5,-4.5){\ensuremath{#4}}
% ellipsis and centre-text
\thicklines \put(-1.5,0){\line(1,0){.1}}
\put(-1.5,-1){\line(1,0){.1}} \put(-1.5,-2){\line(1,0){.1}}
\thinlines \put(2,-0.5){\ensuremath{#1}}
\end{picture}}
%

% Centred transistor, for use in text
%
\newcommand{\ctransistor}[5]
    {\raisebox{-6\volt}{\transistor{#1}{#2}{#3}{#4}{#5}}}

% A brief transistor
%
\newcommand{\bftransistor}[4]
{\setlength{\unitlength}{1\volt}
\begin{picture}(18,12)(-5,-6)
% main triangle
\put(0,6){\line(0,-1){12}} \put(0,-6){\line(3,2){9}}
\put(0,6){\line(3,-2){9}}
% legs
\put(9,0){\line(1,0){2}} \put(-2,4){\line(1,0){2}}
\put(-2,-4){\line(1,0){2}}
% leg-labels
\put(12,-0.5){\ensuremath{#4}} \put(-5,3.5){\ensuremath{#2}}
\put(-5,-4.5){\ensuremath{#3}}
% ellipsis and centre-text
\thicklines \put(-1.5,1){\line(1,0){.1}}
\put(-1.5,0){\line(1,0){.1}} \put(-1.5,-1){\line(1,0){.1}}
\thinlines \put(2,-0.5){\ensuremath{#1}}
\end{picture}}
%

% Composite of transistors
%
\newcommand{\comptrans}[4]
{\setlength{\volt}{.5ex} \setlength{\unitlength}{1\volt}
\begin{picture}(80,72)(0,-36)
\put(0,23){\bftransistor{#1}{}{}{}}
\put(0,6){\bftransistor{#2}{}{}{}}
\put(0,-35){\bftransistor{#3}{}{}{}}
\put(63,-6){\transistor{#4}{}{}{}} \put(17,29){\line(2,-1){50}}
\put(17,12){\line(5,-1){50}} \put(17,-29){\line(2,1){50}}
\thicklines \put(24,-5){\line(1,0){.4}}
\put(24,-9){\line(1,0){.4}} \put(24,-13){\line(1,0){.4}}
\thinlines \setlength{\volt}{1ex}
\end{picture}}

% Centred composite of transistors, for use in text
%
\newcommand{\ccomptrans}[4]
    {\raisebox{-36\volt}{\comptrans{#1}{#2}{#3}{#4}}}

% A 3-legged transistor; used like `threetransistor{a}{s1}{s2}{s3}{s}'
%
\newcommand{\threetransistor}[5]
{\setlength{\unitlength}{1\volt}
\begin{picture}(18,12)(-5,-6)
% main triangle
\put(0,6){\line(0,-1){12}} \put(0,-6){\line(3,2){9}}
\put(0,6){\line(3,-2){9}}
% legs
\put(9,0){\line(1,0){2}} \put(-2,4){\line(1,0){2}}
\put(-2,0){\line(1,0){2}} \put(-2,-4){\line(1,0){2}}
% leg-labels
\put(12,-0.5){\ensuremath{#5}} \put(-5,3.5){\ensuremath{#2}}
\put(-5,-0.5){\ensuremath{#3}} \put(-5,-4.5){\ensuremath{#4}}
% centre-text
\put(2,-0.5){\ensuremath{#1}}
\end{picture}}
%

% A 2-legged transistor; used like `twotransistor{a}{s1}{s2}{s}'
%
\newcommand{\twotransistor}[4]
{\setlength{\unitlength}{1\volt}
\begin{picture}(18,12)(-5,-6)
% main triangle
\put(0,6){\line(0,-1){12}} \put(0,-6){\line(3,2){9}}
\put(0,6){\line(3,-2){9}}
% legs
\put(9,0){\line(1,0){2}} \put(-2,4){\line(1,0){2}}
\put(-2,-4){\line(1,0){2}}
% leg-labels
\put(12,-0.5){\ensuremath{#4}} \put(-5,3.5){\ensuremath{#2}}
\put(-5,-4.5){\ensuremath{#3}}
% centre-text
\put(2,-0.5){\ensuremath{#1}}
\end{picture}}
%

% A no-legged transistor; used like `notransistor{a}{s}'
%
\newcommand{\notransistor}[2]
{\setlength{\unitlength}{1\volt}
\begin{picture}(18,12)(-5,-6)
% main triangle
\put(0,6){\line(0,-1){12}} \put(0,-6){\line(3,2){9}}
\put(0,6){\line(3,-2){9}}
% leg
\put(9,0){\line(1,0){2}}
% leg-labels
\put(12,-0.5){\ensuremath{#2}}
% centre-text
\put(2,-0.5){\ensuremath{#1}}
\end{picture}}
%

% The picture for arrows in the free mon cat on a multicat
%
\newcommand{\freemoncatpic}
{\setlength{\volt}{.6ex} \setlength{\unitlength}{1\volt}
\begin{picture}(18,48)
\put(0,3){\twotransistor{\theta_3}{a_4}{a_5}{a'_3}}
\put(0,18){\notransistor{\theta_2}{a'_2}}
\put(0,33){\threetransistor{\theta_1}{a_1}{a_2}{a_3}{a'_1}}
%   \put(4.2,0){\framebox(12.6,48){}}
\put(5.7,0){\framebox(10.3,48){}}
\end{picture}}
%

%%%%%%%%%%%%%%%%%%%%%%%%%%%%%%%%%%%%%%%%%%%%%%%%%%%%%%%%%%%%%%%%%%%%%%%%%
%
% Miscellaneous pictures

% _+1-multicat picture
%
\newcommand{\dotty}{\raisebox{-.4ex}{\ensuremath{\!\scriptstyle\bullet}}}
\newcommand{\discfibpic}
{\setlength{\unitlength}{1em}       % if you change this, also
                    % change cdiscfib below
\begin{picture}(16,9)(-6,0)
% main oval
\put(4,2){\oval(8,4)}
% dots
\put(4,8){\dotty} \put(2,2){\dotty} \put(4,1){\dotty}
\put(6,1){\dotty} \put(5,3){\dotty}
% arrows
\put(4,8){\vector(-1,-3){2}} \put(2,2){\vector(2,-1){2}}
\put(4,8){\vector(0,-1){7}} \put(2,2){\vector(3,1){3}}
\put(4,1){\vector(1,0){2}}
% text
\put(4.2,8){0} \put(4.2,5.5){\ensuremath{\theta\of y}}
\put(2.5,5.5){\ensuremath{y}} \put(1.2,1.7){\ensuremath{a}}
\put(3.6,.2){\ensuremath{a'}} \put(2.1,.9){\ensuremath{\theta}}
\put(8.5,1.5){\ensuremath{C_0}}
\end{picture}}
%

% Centred version of this, for use in text
%
\newcommand{\cdiscfibpic}{\raisebox{-4.5em}{\discfibpic}}
%

%-----------------------------------------------------------------------------

\makeatletter

\def\diagram{\m@th\leftwidth=\z@ \rightwidth=\z@ \topheight=\z@
\botheight=\z@ \setbox\@picbox\hbox\bgroup}

\def\enddiagram{\egroup\wd\@picbox\rightwidth\unitlength
\ht\@picbox\topheight\unitlength \dp\@picbox\botheight\unitlength
\hskip\leftwidth\unitlength\box\@picbox}

\def\bfig{\begin{diagram}}
\def\efig{\end{diagram}}
\newcount\wideness \newcount\leftwidth \newcount\rightwidth
\newcount\highness \newcount\topheight \newcount\botheight

\def\ratchet#1#2{\ifnum#1<#2 \global #1=#2 \fi}

\def\putbox(#1,#2)#3{%
\horsize{\wideness}{#3} \divide\wideness by 2 {\advance\wideness
by #1 \ratchet{\rightwidth}{\wideness}} {\advance\wideness by -#1
\ratchet{\leftwidth}{\wideness}} \vertsize{\highness}{#3}
\divide\highness by 2 {\advance\highness by #2
\ratchet{\topheight}{\highness}} {\advance\highness by -#2
\ratchet{\botheight}{\highness}} \put(#1,#2){\makebox(0,0){$#3$}}}

\def\putlbox(#1,#2)#3{%
\horsize{\wideness}{#3} {\advance\wideness by #1
\ratchet{\rightwidth}{\wideness}} {\ratchet{\leftwidth}{-#1}}
\vertsize{\highness}{#3} \divide\highness by 2 {\advance\highness
by #2 \ratchet{\topheight}{\highness}} {\advance\highness by -#2
\ratchet{\botheight}{\highness}}
\put(#1,#2){\makebox(0,0)[l]{$#3$}}}

\def\putrbox(#1,#2)#3{%
\horsize{\wideness}{#3} {\ratchet{\rightwidth}{#1}}
{\advance\wideness by -#1 \ratchet{\leftwidth}{\wideness}}
\vertsize{\highness}{#3} \divide\highness by 2 {\advance\highness
by #2 \ratchet{\topheight}{\highness}} {\advance\highness by -#2
\ratchet{\botheight}{\highness}}
\put(#1,#2){\makebox(0,0)[r]{$#3$}}}

\def\adjust[#1]{} % For compatibility

\newcount \coefa
\newcount \coefb
\newcount \coefc
\newcount\tempcounta
\newcount\tempcountb
\newcount\tempcountc
\newcount\tempcountd
\newcount\xext
\newcount\yext
\newcount\xoff
\newcount\yoff
\newcount\gap%
\newcount\arrowtypea
\newcount\arrowtypeb
\newcount\arrowtypec
\newcount\arrowtyped
\newcount\arrowtypee
\newcount\height
\newcount\width
\newcount\xpos
\newcount\ypos
\newcount\run
\newcount\rise
\newcount\arrowlength
\newcount\halflength
\newcount\arrowtype
\newdimen\tempdimen
\newdimen\xlen
\newdimen\ylen
\newsavebox{\tempboxa}%
\newsavebox{\tempboxb}%
\newsavebox{\tempboxc}%

\newdimen\w@dth

\def\setw@dth#1#2{\setbox\z@\hbox{\m@th$#1$}\w@dth=\wd\z@
\setbox\@ne\hbox{\m@th$#2$}\ifnum\w@dth<\wd\@ne \w@dth=\wd\@ne \fi
\advance\w@dth by 1.2em}

%The definitions below look more elaborate than they need to be.
%The reason is that an empty asscript will still cause extra vertical
%spacing and the only way to avoid ugly extra space seems to be using
%some such method as this.

\def\t@^#1_#2{\allowbreak\def\n@one{#1}\def\n@two{#2}\mathrel
{\setw@dth{#1}{#2} \mathop{\hbox to
\w@dth{\rightarrowfill}}\limits \ifx\n@one\empty\else
^{\box\z@}\fi \ifx\n@two\empty\else _{\box\@ne}\fi}}
\def\t@@^#1{\@ifnextchar_{\t@^{#1}}{\t@^{#1}_{}}}
\def\to{\@ifnextchar^{\t@@}{\t@@^{}}}

\def\t@left^#1_#2{\def\n@one{#1}\def\n@two{#2}\mathrel{\setw@dth{#1}{#2}
\mathop{\hbox to \w@dth{\leftarrowfill}}\limits
\ifx\n@one\empty\else ^{\box\z@}\fi \ifx\n@two\empty\else
_{\box\@ne}\fi}}
\def\t@@left^#1{\@ifnextchar_{\t@left^{#1}}{\t@left^{#1}_{}}}
\def\toleft{\@ifnextchar^{\t@@left}{\t@@left^{}}}

\def\two@^#1_#2{\allowbreak
\def\n@one{#1}\def\n@two{#2}\mathrel{\setw@dth{#1}{#2}
\mathop{\vcenter{\lineskip\z@\baselineskip\z@
                 \hbox to \w@dth{\rightarrowfill}%
                 \hbox to \w@dth{\rightarrowfill}}%
       }\limits
\ifx\n@one\empty\else ^{\box\z@}\fi \ifx\n@two\empty\else
_{\box\@ne}\fi}}
\def\tw@@^#1{\@ifnextchar _{\two@^{#1}}{\two@^{#1}_{}}}
\def\two{\@ifnextchar ^{\tw@@}{\tw@@^{}}}

\def\tofr@^#1_#2{\def\n@one{#1}\def\n@two{#2}\mathrel{\setw@dth{#1}{#2}
\mathop{\vcenter{\hbox to \w@dth{\rightarrowfill}\kern-1.7ex
                 \hbox to \w@dth{\leftarrowfill}}%
       }\limits
\ifx\n@one\empty\else ^{\box\z@}\fi \ifx\n@two\empty\else
_{\box\@ne}\fi}}
\def\t@fr@^#1{\@ifnextchar_ {\tofr@^{#1}}{\tofr@^{#1}_{}}}
\def\tofro{\@ifnextchar^ {\t@fr@}{\t@fr@^{}}}

\def\epi{\mathop{\mathchar"221\mkern -12mu\mathchar"221}\limits}
\def\leftepi{\mathop{\mathchar"220\mkern -12mu\mathchar"220}\limits}
\def\mon{\mathop{\m@th\hbox to
      14.6\P@{\lasyb\char'51\hskip-2.1\P@$\arrext$\hss
$\mathord\rightarrow$}}\limits} % width of \epi
\def\leftmono{\mathrel{\m@th\hbox to
14.6\P@{$\mathord\leftarrow$\hss$\arrext$\hskip-2.1\P@\lasyb\char'50%
}}\limits} % width of \epi
\mathchardef\arrext="0200       % amr minus for arrow extension (see \into)

\setlength{\unitlength}{.01em}%
\def\settypes(#1,#2,#3){\arrowtypea#1 \arrowtypeb#2 \arrowtypec#3}
\def\settoheight#1#2{\setbox\@tempboxa\hbox{#2}#1\ht\@tempboxa\relax}%
\def\settodepth#1#2{\setbox\@tempboxa\hbox{#2}#1\dp\@tempboxa\relax}%
\def\settokens`#1`#2`#3`#4`{%
     \def\tokena{#1}\def\tokenb{#2}\def\tokenc{#3}\def\tokend{#4}}
\def\setsqparms[#1`#2`#3`#4;#5`#6]{%
\arrowtypea #1 \arrowtypeb #2 \arrowtypec #3 \arrowtyped #4
\width #5 \height #6 }
\def\setpos(#1,#2){\xpos=#1 \ypos#2}

\def\settriparms[#1`#2`#3;#4]{\settripairparms[#1`#2`#3`1`1;#4]}%

\def\settripairparms[#1`#2`#3`#4`#5;#6]{%
\arrowtypea #1 \arrowtypeb #2 \arrowtypec #3 \arrowtyped #4
\arrowtypee #5 \width #6 \height #6 }

\def\resetparms{\settripairparms[1`1`1`1`1;500]\width 500}%default values%

\resetparms

\def\mvector(#1,#2)#3{%%
\put(0,0){\vector(#1,#2){#3}}%
\put(0,0){\vector(#1,#2){26}}%
}
\def\evector(#1,#2)#3{{%%
\arrowlength #3
\put(0,0){\vector(#1,#2){\arrowlength}}%
\advance \arrowlength by-30
\put(0,0){\vector(#1,#2){\arrowlength}}%
}}

\def\horsize#1#2{%
\settowidth{\tempdimen}{$#2$}%
#1=\tempdimen \divide #1 by\unitlength }

\def\vertsize#1#2{%
\settoheight{\tempdimen}{$#2$}%
#1=\tempdimen
\settodepth{\tempdimen}{$#2$}%
\advance #1 by\tempdimen \divide #1 by\unitlength }

\def\putvector(#1,#2)(#3,#4)#5#6{{%
\ifnum3<\arrowtype \putdashvector(#1,#2)(#3,#4)#5\arrowtype \else
\ifnum\arrowtype<-3 \putdashvector(#1,#2)(#3,#4)#5\arrowtype \else
\xpos=#1 \ypos=#2 \run=#3 \rise=#4 \arrowlength=#5 \ifnum
\arrowtype<0
    \ifnum \run=0
        \advance \ypos by-\arrowlength
    \else
        \tempcounta \arrowlength
        \multiply \tempcounta by\rise
        \divide \tempcounta by\run
        \ifnum\run>0
            \advance \xpos by\arrowlength
            \advance \ypos by\tempcounta
        \else
            \advance \xpos by-\arrowlength
            \advance \ypos by-\tempcounta
        \fi
    \fi
    \multiply \arrowtype by-1
    \multiply \rise by-1
    \multiply \run by-1
\fi \ifcase \arrowtype
\or \put(\xpos,\ypos){\vector(\run,\rise){\arrowlength}}%
\or \put(\xpos,\ypos){\mvector(\run,\rise)\arrowlength}%
\or \put(\xpos,\ypos){\evector(\run,\rise){\arrowlength}}%
\fi\fi\fi }}

\def\putsplitvector(#1,#2)#3#4{%%
\xpos #1 \ypos #2 \arrowtype #4 \halflength #3 \arrowlength #3
\gap 140 \advance \halflength by-\gap \divide \halflength by2
\ifnum\arrowtype>0
   \ifcase \arrowtype
   \or \put(\xpos,\ypos){\line(0,-1){\halflength}}%
       \advance\ypos by-\halflength
       \advance\ypos by-\gap
       \put(\xpos,\ypos){\vector(0,-1){\halflength}}%
   \or \put(\xpos,\ypos){\line(0,-1)\halflength}%
       \put(\xpos,\ypos){\vector(0,-1)3}%
       \advance\ypos by-\halflength
       \advance\ypos by-\gap
       \put(\xpos,\ypos){\vector(0,-1){\halflength}}%
   \or \put(\xpos,\ypos){\line(0,-1)\halflength}%
       \advance\ypos by-\halflength
       \advance\ypos by-\gap
       \put(\xpos,\ypos){\evector(0,-1){\halflength}}%
   \fi
\else \arrowtype=-\arrowtype
   \ifcase\arrowtype
   \or \advance \ypos by-\arrowlength
       \put(\xpos,\ypos){\line(0,1){\halflength}}%
       \advance\ypos by\halflength
       \advance\ypos by\gap
       \put(\xpos,\ypos){\vector(0,1){\halflength}}%
   \or \advance \ypos by-\arrowlength
       \put(\xpos,\ypos){\line(0,1)\halflength}%
       \put(\xpos,\ypos){\vector(0,1)3}%
       \advance\ypos by\halflength
       \advance\ypos by\gap
       \put(\xpos,\ypos){\vector(0,1){\halflength}}%
   \or \advance \ypos by-\arrowlength
       \put(\xpos,\ypos){\line(0,1)\halflength}%
       \advance\ypos by\halflength
       \advance\ypos by\gap
       \put(\xpos,\ypos){\evector(0,1){\halflength}}%
   \fi
\fi }

\def\putmorphism(#1)(#2,#3)[#4`#5`#6]#7#8#9{{%
\run #2 \rise #3 \ifnum\rise=0
  \puthmorphism(#1)[#4`#5`#6]{#7}{#8}#9%
\else\ifnum\run=0
  \putvmorphism(#1)[#4`#5`#6]{#7}{#8}#9%
\else
\setpos(#1)%
\arrowlength #7 \arrowtype #8 \ifnum\run=0 \else\ifnum\rise=0
\else \ifnum\run>0
    \coefa=1
\else
   \coefa=-1
\fi \ifnum\arrowtype>0
   \coefb=0
   \coefc=-1
\else
   \coefb=\coefa
   \coefc=1
   \arrowtype=-\arrowtype
\fi \width=2 \multiply \width by\run \divide \width by\rise
\ifnum \width<0  \width=-\width\fi \advance\width by60 \if l#9
\width=-\width\fi
\putbox(\xpos,\ypos){#4}%            %node 1
{\multiply \coefa by\arrowlength%      %node 2
\advance\xpos by\coefa \multiply \coefa by\rise \divide \coefa
by\run \advance \ypos by\coefa
\putbox(\xpos,\ypos){#5} }%
{\multiply \coefa by\arrowlength%      %label
\divide \coefa by2 \advance \xpos by\coefa \advance \xpos by\width
\multiply \coefa by\rise \divide \coefa by\run \advance \ypos
by\coefa
\if l#9%
   \putrbox(\xpos,\ypos){#6}%
\else\if r#9%
   \putlbox(\xpos,\ypos){#6}%
\fi\fi }%
{\multiply \rise by-\coefc%             %arrow
\multiply \run by-\coefc \multiply \coefb by\arrowlength \advance
\xpos by\coefb \multiply \coefb by\rise \divide \coefb by\run
\advance \ypos by\coefb \multiply \coefc by70 \advance \ypos
by\coefc \multiply \coefc by\run \divide \coefc by\rise \advance
\xpos by\coefc \multiply \coefa by140 \multiply \coefa by\run
\divide \coefa by\rise \advance \arrowlength by\coefa
\ifcase\arrowtype
\or \put(\xpos,\ypos){\vector(\run,\rise){\arrowlength}}%
\or \put(\xpos,\ypos){\mvector(\run,\rise){\arrowlength}}%
\or \put(\xpos,\ypos){\evector(\run,\rise){\arrowlength}}%
\fi}\fi\fi\fi\fi}}

\newcount\numbdashes \newcount\lengthdash \newcount\increment

\def\howmanydashes{% Actually returns both number and length
\numbdashes=\arrowlength \lengthdash=40 \divide\numbdashes by
\lengthdash \lengthdash=\arrowlength \divide\lengthdash by
\numbdashes
%This futzing around is to minimize round-off error.
\increment=\lengthdash \multiply\lengthdash by 3
\divide\lengthdash by 5 }

\def\putdashvector(#1)(#2,#3)#4#5{%
\ifnum#3=0 \putdashhvector(#1){#4}#5 \else \ifnum#2=0
\putdashvvector(#1){#4}#5\fi\fi}

\def\putdashhvector(#1,#2)#3#4{{%
\arrowlength=#3 \howmanydashes
\multiput(#1,#2)(\increment,0){\numbdashes}%
{\vrule height .4pt width \lengthdash\unitlength} \arrowtype=#4
\xpos=#1 \ifnum\arrowtype<0 \advance\arrowtype by 7 \fi
\ifcase\arrowtype \or \advance\xpos by 10
    \put(\xpos,#2){\vector(-1,0){\lengthdash}}
    \advance\xpos by 40
    \put(\xpos,#2){\vector(-1,0){\lengthdash}}
\or \advance \xpos by 10
    \put(\xpos,#2){\vector(-1,0){\lengthdash}}
    \advance\xpos by  \arrowlength
    \advance\xpos by  -50
    \put(\xpos,#2){\vector(-1,0){\lengthdash}}
\or \advance\xpos by 10
    \put(\xpos,#2){\vector(-1,0){\lengthdash}}
\or \advance\xpos by \arrowlength
    \advance\xpos by -\lengthdash
    \put(\xpos,#2){\vector(1,0){\lengthdash}}
\or {\advance\xpos by 10
    \put(\xpos,#2){\vector(1,0){\lengthdash}}}
    \advance\xpos by \arrowlength
    \advance\xpos by -\lengthdash
    \put(\xpos,#2){\vector(1,0){\lengthdash}}
\or \advance\xpos by \arrowlength
    \advance\xpos by -\lengthdash
    \put(\xpos,#2){\vector(1,0){\lengthdash}}
    \advance\xpos by -40
    \put(\xpos,#2){\vector(1,0){\lengthdash}}
   \fi
}}

\def\putdashvvector(#1,#2)#3#4{{%
\arrowlength=#3 \howmanydashes \ypos=#2 \advance\ypos by
-\arrowlength
\multiput(#1,#2)(0,\increment){\numbdashes}%
    {\vrule width .4pt height \lengthdash\unitlength}
\arrowtype=#4 \ypos=#2 \ifnum\arrowtype<0 \advance\arrowtype by 7
\fi \ifcase\arrowtype \or \advance\ypos by \arrowlength
\advance\ypos by -40
    \put(#1,\ypos){\vector(0,1){\lengthdash}}
    \advance\ypos by -40
    \put(#1,\ypos){\vector(0,1){\lengthdash}}
\or \advance\ypos by 10
    \put(#1,\ypos){\vector(0,1){\lengthdash}}
    \advance\ypos by \arrowlength \advance\ypos by -40
    \put(#1,\ypos){\vector(0,1){\lengthdash}}
\or \advance\ypos by \arrowlength \advance\ypos by -40
    \put(#1,\ypos){\vector(0,1){\lengthdash}}
\or \advance\ypos by 10
    \put(#1,\ypos){\vector(0,-1){\lengthdash}}
\or \advance\ypos by 10
    \put(#1,\ypos){\vector(0,-1){\lengthdash}}
    \advance\ypos by \arrowlength \advance\ypos by -40
    \put(#1,\ypos){\vector(0,-1){\lengthdash}}
\or \advance\ypos by 10
    \put(#1,\ypos){\vector(0,-1){\lengthdash}}
    \advance\ypos by 40
    \put(#1,\ypos){\vector(0,-1){\lengthdash}}
\fi }}

\def\puthmorphism(#1,#2)[#3`#4`#5]#6#7#8{{%
\xpos #1 \ypos #2 \width #6 \arrowlength #6 \arrowtype=#7
\putbox(\xpos,\ypos){#3\vphantom{#4}}%
{\advance \xpos by\arrowlength
\putbox(\xpos,\ypos){\vphantom{#3}#4}}%
\horsize{\tempcounta}{#3}%
\horsize{\tempcountb}{#4}%
\divide \tempcounta by2 \divide \tempcountb by2 \advance
\tempcounta by30 \advance \tempcountb by30 \advance \xpos
by\tempcounta \advance \arrowlength by-\tempcounta \advance
\arrowlength by-\tempcountb
\putvector(\xpos,\ypos)(1,0)\arrowlength\arrowtype \divide
\arrowlength by2 \advance \xpos by\arrowlength
\vertsize{\tempcounta}{#5}%
\divide\tempcounta by2 \advance \tempcounta by20
\if a#8 %
   \advance \ypos by\tempcounta
   \putbox(\xpos,\ypos){#5}%
\else
   \advance \ypos by-\tempcounta
   \putbox(\xpos,\ypos){#5}%
\fi}}

\def\putvmorphism(#1,#2)[#3`#4`#5]#6#7#8{{%
\xpos #1 \ypos #2 \arrowlength #6 \arrowtype #7
\settowidth{\xlen}{$#5$}%
\putbox(\xpos,\ypos){#3}%
{\advance \ypos by-\arrowlength
\putbox(\xpos,\ypos){#4}}%
{\advance\arrowlength by-140 \advance \ypos by-70 \ifdim\xlen>0pt
   \if m#8%
      \putsplitvector(\xpos,\ypos)\arrowlength\arrowtype
   \else
   \putvector(\xpos,\ypos)(0,-1)\arrowlength\arrowtype
   \fi
\else
   \putvector(\xpos,\ypos)(0,-1)\arrowlength\arrowtype
\fi}%
\ifdim\xlen>0pt
   \divide \arrowlength by2
   \advance\ypos by-\arrowlength
   \if l#8%
      \advance \xpos by-40
      \putrbox(\xpos,\ypos){#5}%
   \else\if r#8%
      \advance \xpos by40
      \putlbox(\xpos,\ypos){#5}%
   \else
      \putbox(\xpos,\ypos){#5}%
   \fi\fi
\fi }}

\def\putsquarep<#1>(#2)[#3;#4`#5`#6`#7]{{%
\setsqparms[#1]%
\setpos(#2)%
\settokens`#3`%
\puthmorphism(\xpos,\ypos)[\tokenc`\tokend`{#7}]{\width}{\arrowtyped}b%
\advance\ypos by \height
\puthmorphism(\xpos,\ypos)[\tokena`\tokenb`{#4}]{\width}{\arrowtypea}a%
\putvmorphism(\xpos,\ypos)[``{#5}]{\height}{\arrowtypeb}l%
\advance\xpos by \width
\putvmorphism(\xpos,\ypos)[``{#6}]{\height}{\arrowtypec}r%
}}

\def\putsquare{\@ifnextchar <{\putsquarep}{\putsquarep%
   <\arrowtypea`\arrowtypeb`\arrowtypec`\arrowtyped;\width`\height>}}
\def\square{\@ifnextchar< {\squarep}{\squarep
   <\arrowtypea`\arrowtypeb`\arrowtypec`\arrowtyped;\width`\height>}}
                                                   %         #6
\def\squarep<#1>[#2`#3`#4`#5;#6`#7`#8`#9]{{%       %     #2------>#3
\setsqparms[#1]%                                   %      |       |
\diagram%                                          %      |       |
\putsquarep<\arrowtypea`\arrowtypeb`\arrowtypec`%  %    #7|       |#8
\arrowtyped;\width`\height>%                       %      |       |
(0,0)[#2`#3`#4`{#5};#6`#7`#8`{#9}]%                %      |       |
\enddiagram%                                       %      v       v
}}                                                 %     #4------>#5
                                                   %         #9
\def\putptrianglep<#1>(#2,#3)[#4`#5`#6;#7`#8`#9]{{%
\settriparms[#1]%
\xpos=#2 \ypos=#3 \advance\ypos by \height
\puthmorphism(\xpos,\ypos)[#4`#5`{#7}]{\height}{\arrowtypea}a%
\putvmorphism(\xpos,\ypos)[`#6`{#8}]{\height}{\arrowtypeb}l%
\advance\xpos by\height
\putmorphism(\xpos,\ypos)(-1,-1)[``{#9}]{\height}{\arrowtypec}r%
}}

\def\putptriangle{\@ifnextchar <{\putptrianglep}{\putptrianglep
   <\arrowtypea`\arrowtypeb`\arrowtypec;\height>}}
\def\ptriangle{\@ifnextchar <{\ptrianglep}{\ptrianglep
   <\arrowtypea`\arrowtypeb`\arrowtypec;\height>}}
                                              %          #5
\def\ptrianglep<#1>[#2`#3`#4;#5`#6`#7]{{%%    %      #2----->#3
\settriparms[#1]%                             %      |      /
\diagram%                                     %      |     /
\putptrianglep<\arrowtypea`\arrowtypeb`%      %    #6|    /#7
\arrowtypec;\height>%                         %      |   /
(0,0)[#2`#3`#4;#5`#6`{#7}]%                   %      |  /
\enddiagram%%                                 %      v v
}}                                            %      #4

\def\putqtrianglep<#1>(#2,#3)[#4`#5`#6;#7`#8`#9]{{%
\settriparms[#1]%
\xpos=#2 \ypos=#3 \advance\ypos by\height
\puthmorphism(\xpos,\ypos)[#4`#5`{#7}]{\height}{\arrowtypea}a%
\putmorphism(\xpos,\ypos)(1,-1)[``{#8}]{\height}{\arrowtypeb}l%
\advance\xpos by\height
\putvmorphism(\xpos,\ypos)[`#6`{#9}]{\height}{\arrowtypec}r%
}}

\def\putqtriangle{\@ifnextchar <{\putqtrianglep}{\putqtrianglep
   <\arrowtypea`\arrowtypeb`\arrowtypec;\height>}}
\def\qtriangle{\@ifnextchar <{\qtrianglep}{\qtrianglep
   <\arrowtypea`\arrowtypeb`\arrowtypec;\height>}}
                                              %           #5
\def\qtrianglep<#1>[#2`#3`#4;#5`#6`#7]{{%%    %        #2----->#3
\settriparms[#1]%                             %         \      |
\width=\height                                %          \     |
\diagram%                                     %         #6\    |#7
\putqtrianglep<\arrowtypea`\arrowtypeb`%      %            \   |
\arrowtypec;\height>%                         %             \  |
(0,0)[#2`#3`#4;#5`#6`{#7}]%                   %              v v
\enddiagram%%                                 %               #4
}}

\def\putdtrianglep<#1>(#2,#3)[#4`#5`#6;#7`#8`#9]{{%
\settriparms[#1]%
\xpos=#2 \ypos=#3
\puthmorphism(\xpos,\ypos)[#5`#6`{#9}]{\height}{\arrowtypec}b%
\advance\xpos by \height \advance\ypos by\height
\putmorphism(\xpos,\ypos)(-1,-1)[``{#7}]{\height}{\arrowtypea}l%
\putvmorphism(\xpos,\ypos)[#4``{#8}]{\height}{\arrowtypeb}r%
}}

\def\putdtriangle{\@ifnextchar <{\putdtrianglep}{\putdtrianglep
   <\arrowtypea`\arrowtypeb`\arrowtypec;\height>}}
\def\dtriangle{\@ifnextchar <{\dtrianglep}{\dtrianglep
   <\arrowtypea`\arrowtypeb`\arrowtypec;\height>}}
                                              %                   #2
\def\dtrianglep<#1>[#2`#3`#4;#5`#6`#7]{{%%    %                  / |
\settriparms[#1]%                             %                 /  |
\width=\height                                %              #5/   |#6
\diagram%                                     %               /    |
\putdtrianglep<\arrowtypea`\arrowtypeb`%      %              /     |
\arrowtypec;\height>%                         %             v      v
(0,0)[#2`#3`#4;#5`#6`{#7}]%                   %            #3----->#4
\enddiagram%%                                 %                #7
}}

\def\putbtrianglep<#1>(#2,#3)[#4`#5`#6;#7`#8`#9]{{%
\settriparms[#1]%
\xpos=#2 \ypos=#3
\puthmorphism(\xpos,\ypos)[#5`#6`{#9}]{\height}{\arrowtypec}b%
\advance\ypos by\height
\putmorphism(\xpos,\ypos)(1,-1)[``{#8}]{\height}{\arrowtypeb}r%
\putvmorphism(\xpos,\ypos)[#4``{#7}]{\height}{\arrowtypea}l%
}}

\def\putbtriangle{\@ifnextchar <{\putbtrianglep}{\putbtrianglep
   <\arrowtypea`\arrowtypeb`\arrowtypec;\height>}}
\def\btriangle{\@ifnextchar <{\btrianglep}{\btrianglep
   <\arrowtypea`\arrowtypeb`\arrowtypec;\height>}}
                                             %              #2
\def\btrianglep<#1>[#2`#3`#4;#5`#6`#7]{{%%   %              | \
\settriparms[#1]%                            %              |  \
\width=\height                               %            #5|   \#6
\diagram%                                    %              |    \
\putbtrianglep<\arrowtypea`\arrowtypeb`%     %              |     \
\arrowtypec;\height>%                        %              v      v
(0,0)[#2`#3`#4;#5`#6`{#7}]%                  %              #3----->#4
\enddiagram%%                                %                 #7
}}

\def\putAtrianglep<#1>(#2,#3)[#4`#5`#6;#7`#8`#9]{{%
\settriparms[#1]%
\xpos=#2 \ypos=#3 {\multiply \height by2
\puthmorphism(\xpos,\ypos)[#5`#6`{#9}]{\height}{\arrowtypec}b}%
\advance\xpos by\height \advance\ypos by\height
\putmorphism(\xpos,\ypos)(-1,-1)[#4``{#7}]{\height}{\arrowtypea}l%
\putmorphism(\xpos,\ypos)(1,-1)[``{#8}]{\height}{\arrowtypeb}r%
}}

\def\putAtriangle{\@ifnextchar <{\putAtrianglep}{\putAtrianglep
   <\arrowtypea`\arrowtypeb`\arrowtypec;\height>}}
\def\Atriangle{\@ifnextchar <{\Atrianglep}{\Atrianglep
   <\arrowtypea`\arrowtypeb`\arrowtypec;\height>}}
                                                   %           #2
\def\Atrianglep<#1>[#2`#3`#4;#5`#6`#7]{{%%         %         /   \
\settriparms[#1]%                                  %        /     \
\width=\height                                     %     #5/       \#6
\diagram%                                          %      /         \
\putAtrianglep<\arrowtypea`\arrowtypeb`%           %     /           \
\arrowtypec;\height>%                              %    v             v
(0,0)[#2`#3`#4;#5`#6`{#7}]%                        %   #3------------>#4
\enddiagram%%                                      %          #7
}}

\def\putAtrianglepairp<#1>(#2)[#3;#4`#5`#6`#7`#8]{{%
\settripairparms[#1]%
\setpos(#2)%
\settokens`#3`%
\puthmorphism(\xpos,\ypos)[\tokenb`\tokenc`{#7}]{\height}{\arrowtyped}b%
\advance\xpos by\height
\puthmorphism(\xpos,\ypos)[\phantom{\tokenc}`\tokend`{#8}]%
{\height}{\arrowtypee}b%
\advance\ypos by\height
\putmorphism(\xpos,\ypos)(-1,-1)[\tokena``{#4}]{\height}{\arrowtypea}l%
\putvmorphism(\xpos,\ypos)[``{#5}]{\height}{\arrowtypeb}m%
\putmorphism(\xpos,\ypos)(1,-1)[``{#6}]{\height}{\arrowtypec}r%
}}

\def\putAtrianglepair{\@ifnextchar <{\putAtrianglepairp}{\putAtrianglepairp%
   <\arrowtypea`\arrowtypeb`\arrowtypec`\arrowtyped`\arrowtypee;\height>}}
\def\Atrianglepair{\@ifnextchar <{\Atrianglepairp}{\Atrianglepairp%
   <\arrowtypea`\arrowtypeb`\arrowtypec`\arrowtyped`\arrowtypee;\height>}}

\def\Atrianglepairp<#1>[#2;#3`#4`#5`#6`#7]{{%           %  #2a
\settripairparms[#1]%                         %           / | \
\settokens`#2`%                               %          /  |  \
\width=\height                                %       #3/  #4   \#5
\diagram%                                     %        /    |    \
\putAtrianglepairp                            %       /     |     \
<\arrowtypea`\arrowtypeb`\arrowtypec`%        %      v      v      v
\arrowtyped`\arrowtypee;\height>%             %     #2b---->#2c---->#2d
(0,0)[{#2};#3`#4`#5`#6`{#7}]%                 %         #6     #7
\enddiagram%%
}}

\def\putVtrianglep<#1>(#2,#3)[#4`#5`#6;#7`#8`#9]{{%
\settriparms[#1]%
\xpos=#2 \ypos=#3 \advance\ypos by\height {\multiply\height by2
\puthmorphism(\xpos,\ypos)[#4`#5`{#7}]{\height}{\arrowtypea}a}%
\putmorphism(\xpos,\ypos)(1,-1)[`#6`{#8}]{\height}{\arrowtypeb}l%
\advance\xpos by\height \advance\xpos by\height
\putmorphism(\xpos,\ypos)(-1,-1)[``{#9}]{\height}{\arrowtypec}r%
}}

\def\putVtriangle{\@ifnextchar <{\putVtrianglep}{\putVtrianglep
   <\arrowtypea`\arrowtypeb`\arrowtypec;\height>}}
\def\Vtriangle{\@ifnextchar <{\Vtrianglep}{\Vtrianglep
   <\arrowtypea`\arrowtypeb`\arrowtypec;\height>}}
                                               %               #5
\def\Vtrianglep<#1>[#2`#3`#4;#5`#6`#7]{{%%     %        #2------------->#3
\settriparms[#1]%                              %         \             /
\width=\height                                 %          \           /
\diagram%                                      %         #6\         /#7
\putVtrianglep<\arrowtypea`\arrowtypeb`%       %            \       /
\arrowtypec;\height>%                          %             \     /
(0,0)[#2`#3`#4;#5`#6`{#7}]%                    %              v   v
\enddiagram%%                                  %               #4
}}

\def\putVtrianglepairp<#1>(#2)[#3;#4`#5`#6`#7`#8]{{
\settripairparms[#1]%
\setpos(#2)%
\settokens`#3`%
\advance\ypos by\height
\putmorphism(\xpos,\ypos)(1,-1)[`\tokend`{#6}]{\height}{\arrowtypec}l%
\puthmorphism(\xpos,\ypos)[\tokena`\tokenb`{#4}]{\height}{\arrowtypea}a%
\advance\xpos by\height
\puthmorphism(\xpos,\ypos)[\phantom{\tokenb}`\tokenc`{#5}]%
{\height}{\arrowtypeb}a%
\putvmorphism(\xpos,\ypos)[``{#7}]{\height}{\arrowtyped}m%
\advance\xpos by\height
\putmorphism(\xpos,\ypos)(-1,-1)[``{#8}]{\height}{\arrowtypee}r%
}}

\def\putVtrianglepair{\@ifnextchar <{\putVtrianglepairp}{\putVtrianglepairp%
    <\arrowtypea`\arrowtypeb`\arrowtypec`\arrowtyped`\arrowtypee;\height>}}
\def\Vtrianglepair{\@ifnextchar <{\Vtrianglepairp}{\Vtrianglepairp%
    <\arrowtypea`\arrowtypeb`\arrowtypec`\arrowtyped`\arrowtypee;\height>}}
                                               %     #3      #4
\def\Vtrianglepairp<#1>[#2;#3`#4`#5`#6`#7]{{%  %  #2a---->#2b---->#2c
\settripairparms[#1]%                          %   \      |      /
\settokens`#2`%                                %    \     |     /
\diagram%                                      %   #5\   #6    /#7
\putVtrianglepairp                             %      \   |   /
<\arrowtypea`\arrowtypeb`\arrowtypec`%         %       \  |  /
\arrowtyped`\arrowtypee;\height>%              %        v v v
(0,0)[{#2};#3`#4`#5`#6`{#7}]%                  %         #2d
\enddiagram%%
}}

\def\putCtrianglep<#1>(#2,#3)[#4`#5`#6;#7`#8`#9]{{%
\settriparms[#1]%
\xpos=#2 \ypos=#3 \advance\ypos by\height
\putmorphism(\xpos,\ypos)(1,-1)[``{#9}]{\height}{\arrowtypec}l%
\advance\xpos by\height \advance\ypos by\height
\putmorphism(\xpos,\ypos)(-1,-1)[#4`#5`{#7}]{\height}{\arrowtypea}l%
{\multiply\height by 2
\putvmorphism(\xpos,\ypos)[`#6`{#8}]{\height}{\arrowtypeb}r}%
}}

\def\putCtriangle{\@ifnextchar <{\putCtrianglep}{\putCtrianglep
    <\arrowtypea`\arrowtypeb`\arrowtypec;\height>}}
\def\Ctriangle{\@ifnextchar <{\Ctrianglep}{\Ctrianglep
    <\arrowtypea`\arrowtypeb`\arrowtypec;\height>}}
                                             %                 #2
\def\Ctrianglep<#1>[#2`#3`#4;#5`#6`#7]{{%%   %                / |
\settriparms[#1]%                            %             #5/  |
\width=\height                               %              /   |
\diagram%                                    %             v    |
\putCtrianglep<\arrowtypea`\arrowtypeb`%     %           #3     |#6
\arrowtypec;\height>%                        %             \    |
(0,0)[#2`#3`#4;#5`#6`{#7}]%                  %            #7\   |
\enddiagram%%                                %               \  |
}}                                           %                v v
                                             %                 #4
\def\putDtrianglep<#1>(#2,#3)[#4`#5`#6;#7`#8`#9]{{%
\settriparms[#1]%
\xpos=#2 \ypos=#3 \advance\xpos by\height \advance\ypos by\height
\putmorphism(\xpos,\ypos)(-1,-1)[``{#9}]{\height}{\arrowtypec}r%
\advance\xpos by-\height \advance\ypos by\height
\putmorphism(\xpos,\ypos)(1,-1)[`#5`{#8}]{\height}{\arrowtypeb}r%
{\multiply\height by 2
\putvmorphism(\xpos,\ypos)[#4`#6`{#7}]{\height}{\arrowtypea}l}%
}}

\def\putDtriangle{\@ifnextchar <{\putDtrianglep}{\putDtrianglep
    <\arrowtypea`\arrowtypeb`\arrowtypec;\height>}}
\def\Dtriangle{\@ifnextchar <{\Dtrianglep}{\Dtrianglep
   <\arrowtypea`\arrowtypeb`\arrowtypec;\height>}}
                                            %          #2
\def\Dtrianglep<#1>[#2`#3`#4;#5`#6`#7]{{%%  %          | \
\settriparms[#1]%                           %          |  \#6
\width=\height                              %          |   \
\diagram%                                   %          |    v
\putDtrianglep<\arrowtypea`\arrowtypeb`%    %        #5|    #3
\arrowtypec;\height>%                       %          |    /
(0,0)[#2`#3`#4;#5`#6`{#7}]%                 %          |   /#7
\enddiagram%%                               %          |  /
}}                                          %          v v
                                            %          #4
\def\setrecparms[#1`#2]{\width=#1 \height=#2}%
%              #4
%        #3b<-------#3a x #3b
%       ^ |             |
%      /  |             |
%   #5/   |             |
%    /    |             |
%   /     |             |
%  /      |             |
% #3c     |#6           |#3a x #5
%  \      |             |
%   \     |             |
%  #8\    |             |
%     \   |             |
%      \  |             |
%       v v             v
%        #3d<-------#3a x #3d
%              #8

\def\recursep<#1`#2>[#3;#4`#5`#6`#7`#8]{{\m@th
\width=#1 \height=#2 \settokens`#3`
\settowidth{\tempdimen}{$\tokena$} \ifdim\tempdimen=0pt
  \savebox{\tempboxa}{\hbox{$\tokenb$}}%
  \savebox{\tempboxb}{\hbox{$\tokend$}}%
  \savebox{\tempboxc}{\hbox{$#6$}}%
\else
  \savebox{\tempboxa}{\hbox{$\hbox{$\tokena$}\times\hbox{$\tokenb$}$}}%
  \savebox{\tempboxb}{\hbox{$\hbox{$\tokena$}\times\hbox{$\tokend$}$}}%
  \savebox{\tempboxc}{\hbox{$\hbox{$\tokena$}\times\hbox{$#6$}$}}%
\fi \ypos=\height \divide\ypos by 2 \xpos=\ypos \advance\xpos by
\width \bfig
\putCtrianglep<-1`1`1;\ypos>(0,0)[`\tokenc`;#5`#6`{#7}]%
\puthmorphism(\ypos,0)[\tokend`\usebox{\tempboxb}`{#8}]{\width}{-1}b%
\puthmorphism(\ypos,\height)[\tokenb`\usebox{\tempboxa}`{#4}]{\width}{-1}a%
\advance\ypos by \width
\putvmorphism(\ypos,\height)[``\usebox{\tempboxc}]{\height}1r%
\efig }}

\def\recurse{\@ifnextchar <{\recursep}{\recursep<\width`\height>}}

\def\puttwohmorphisms(#1,#2)[#3`#4;#5`#6]#7#8#9{{%
% 1 and 2 are position, 3 and 4 are the nodes, 5 and 6 the labels,
% 7 the distance between node centers and 8 & 9 are the arrow types.
%         #5
% #3 ===========> #4
%         #6
%
\puthmorphism(#1,#2)[#3`#4`]{#7}0a \ypos=#2 \advance\ypos by 20
\puthmorphism(#1,\ypos)[\phantom{#3}`\phantom{#4}`#5]{#7}{#8}a
\advance\ypos by -40
\puthmorphism(#1,\ypos)[\phantom{#3}`\phantom{#4}`#6]{#7}{#9}b }}

\def\puttwovmorphisms(#1,#2)[#3`#4;#5`#6]#7#8#9{{%
% 1 and 2 are position, 3 and 4 are the nodes, 5 and 6 the labels,
% 7 the distance between node centers and 8 & 9 are the arrow types.
%
%              #3
%              ||
%              ||
%           #5 || #6
%              ||
%              ||
%              vv
%              #4
%
\putvmorphism(#1,#2)[#3`#4`]{#7}0a \xpos=#1 \advance\xpos by -20
\putvmorphism(\xpos,#2)[\phantom{#3}`\phantom{#4}`#5]{#7}{#8}l
\advance\xpos by 40
\putvmorphism(\xpos,#2)[\phantom{#3}`\phantom{#4}`#6]{#7}{#9}r }}

\def\puthcoequalizer(#1)[#2`#3`#4;#5`#6`#7]#8#9{{%
% #1 is (\xpos,\ypos), the next 6 are the nodes and arrow labels
% #8 is the distance between each pair of nodes and #9 is the pos of #7
% either a (above) or b (below)
%         #5            #7
% #2 ===========> #3 --------> #4
%         #6
%
\setpos(#1)%
\puttwohmorphisms(\xpos,\ypos)[#2`#3;#5`#6]{#8}11%
\advance\xpos by #8
\puthmorphism(\xpos,\ypos)[\phantom{#3}`#4`#7]{#8}1{#9} }}

\def\putvcoequalizer(#1)[#2`#3`#4;#5`#6`#7]#8#9{{%
% #1 is (\xpos,\ypos), the next 6 are the nodes and arrow labels
% #8 is the distance between each pair of nodes and #9 is the pos of #7
% either l (left) or r (right)
%
%              #2
%              | |
%              | |
%           #5 | | #6
%              | |
%              | |
%              v v
%              #3
%               |
%               |
%            #7 |
%               |
%               v
%              #4
%
\setpos(#1)%
\puttwovmorphisms(\xpos,\ypos)[#2`#3;#5`#6]{#8}11%
\advance\ypos by -#8
\putvmorphism(\xpos,\ypos)[\phantom{#3}`#4`#7]{#8}1{#9} }}

\def\putthreehmorphisms(#1)[#2`#3;#4`#5`#6]#7(#8)#9{{%
% Use: \putthreehmorphisms(xpos,ypos)[lnode`rnode;toplabel`midlabel%
% botlabel]{distance}(toparrowtype,midarrowtype,botarrowtype){position}
\setpos(#1) \settypes(#8)
\if a#9 %
     \vertsize{\tempcounta}{#5}%
     \vertsize{\tempcountb}{#6}%
     \ifnum \tempcounta<\tempcountb \tempcounta=\tempcountb \fi
\else
     \vertsize{\tempcounta}{#4}%
     \vertsize{\tempcountb}{#5}%
     \ifnum \tempcounta<\tempcountb \tempcounta=\tempcountb \fi
\fi \advance \tempcounta by 60
\puthmorphism(\xpos,\ypos)[#2`#3`#5]{#7}{\arrowtypeb}{#9}
\advance\ypos by \tempcounta
\puthmorphism(\xpos,\ypos)[\phantom{#2}`\phantom{#3}`#4]{#7}{\arrowtypea}{#9}
\advance\ypos by -\tempcounta \advance\ypos by -\tempcounta
\puthmorphism(\xpos,\ypos)[\phantom{#2}`\phantom{#3}`#6]{#7}{\arrowtypec}{#9}
}}

\def\setarrowtoks[#1`#2`#3`#4`#5`#6]{%
\def\toka{#1}
\def\tokb{#2}
\def\tokc{#3}
\def\tokd{#4}
\def\toke{#5}
\def\tokf{#6}
}
\def\hex{\@ifnextchar <{\hexp}{\hexp<1000`400>}}
\def\hexp<#1`#2>[#3`#4`#5`#6`#7`#8;#9]{%
\setarrowtoks[#9] \yext=#2 \advance \yext by #2 \xext=#1
\advance\xext by \yext \bfig
\putCtriangle<-1`0`1;#2>(0,0)[`#5`;\tokb``\tokd] \xext=#1
\yext=#2 \advance \yext by #2
\putsquare<1`0`0`1;\xext`\yext>(#2,0)[#3`#4`#7`#8;\toka```\tokf]
\advance \xext by #2
\putDtriangle<0`1`-1;#2>(\xext,0)[`#6`;`\tokc`\toke] \efig }

\makeatother